\documentclass[aps,pre,epsf,preprint]{revtex4}
\usepackage{graphicx}
\usepackage{dcolumn}
\usepackage{bm,multirow}
\usepackage{subfigure}
\usepackage{amssymb}
\usepackage{amsmath}
\usepackage{graphics}
\usepackage{epsfig}
\usepackage{threeparttable}

\begin{document}

\title{Metastable behavior of the spin-$s$ Ising and Blume-Capel ferromagnets: 
A Monte Carlo study}

\author{Moumita Naskar}

\author{Muktish Acharyya}\email{muktish.physics@presiuniv.ac.in}

\affiliation{Department of Physics, Presidency University, 86/1 College 
Street, Kolkata-700073, India}

\author{Erol Vatansever}

\affiliation{Department of Physics, Dokuz Eyl\"{u}l University, TR-35160 
Izmir, Turkey}

\affiliation{Centre for Fluid and Complex Systems, Coventry University, 
Coventry CV1 5FB, United Kingdom}

\author{Nikolaos G. Fytas}\email{nikolaos.fytas@coventry.ac.uk}

\affiliation{Centre for Fluid and Complex Systems, Coventry University, 
Coventry CV1 5FB, United Kingdom}

\date{\today}

\begin{abstract}
	
We present an extensive Monte Carlo investigation of the metastable lifetime through the 
reversal of the magnetization of spin-$s$ Ising and Blume-Capel models, where $s=\{1/2, 1, 3/2,2,5/2,3,7/2\}$. The mean metastable lifetime (or 
reversal time) is studied as a function of the applied magnetic field 
and for both models is found to obey the Becker-D{\"o}ring theory, as was initially 
developed for the case of $s=1/2$ Ising ferromagnet within the classical nucleation theory. Moreover, the decay of metastable volume fraction nicely follows the Avrami's law for all values of $s$ and for both models considered. 
\end{abstract}

\maketitle

\section{Introduction}
\label{introduction}

The metastable behavior of a ferromagnet is an interesting 
field of modern research~\cite{gunton}. It is well-known today that the lifetime of a metastable state plays an important role in the technological area of magnetic 
recording~\cite{techno}. The whole problem dates back to 1935, where the classical theory of
nucleation was developed by Becker and D{\"o}ring~\cite{becker}, predicting
the growth of supercritical droplets. These predictions of the different regimes of such growth depending on the magnitude of the applied magnetic field were successfully verified by extensive Monte Carlo simulation~\cite{stauffer} in Ising ferromagnets. Needless to say that the fruit-fly model of Statistical Physics, the Ising model, is a prototype system that allows a clear investigation of all these phenomena. 

In more recent years, several interesting complications around this fundamental problem have been proposed and clarified. In particular, the effects of a random magnetic field were investigated in Ref.~\cite{naskar1}, where the cut-off value of the random field was predicted to deviate from 
the Becker-D{\"o}ring theory. The role of a uniform anisotropy in  
metastability of the spin-$1$ Blume-Capel ferromagnet was studied in Ref.~\cite{naskar2} and a scaling law relating to the stable magnetization, applied field and anisotropy 
was proposed. Also, the decay of metastable state was found to follow Avrami's law~\cite{avrami}. Finally, the spatial variation of the anisotropy (graded- and step-like over the lattice) to tune the switching time has been proposed in Ref.~\cite{naskar3}, again regarding the anisotropic Blume-Capel ferromagnet.
 
All studies mentioned above focused on the standard spin-$1/2$ Ising and spin-$1$ Blume-Capel ferromagnets. This is of no surprise as the Becker-D{\"o}ring theory was originally proposed for spin-$1/2$ ferromagnets. In the current paper we address the obvious next-step question of whether the predictions of classical nucleation theory can be observed at any spin-$s$ Ising-type ferromagnet. This open problem of understanding reversal processes in magnets with high spin values is not only of great theoretical interest but also intimately connected to the development of modern technologies which are based on controlled switching of the spin state~\cite{shankar}. To this end, we perform in the current paper a first extensive numerical study of the general spin-$s$ Ising and Blume-Capel models for various values of $s$. We investigate the underlying metastable behavior within the Becker-D{\"o}ring theoretical framework, as well the Avrami's law. 
 
The remainder of this manuscript is organized as follows: In the next Section we briefly review the Becker-D{\"o}ring theory and in Sec.~\ref{model} 
we introduce the models and numerical scheme. In Sec.~\ref{results} we present our results and scaling analysis. Finally, we end this contribution with a summary in Sec.~\ref{summary}.

\section{An Overview of Classical Nucleation and Droplet Theory}
\label{theory}

Below the critical temperature $T_{\rm c}$, if a very weak magnetic field (antiparallel to the initial spin direction) is applied to a ferromagnet, 
the system passes through a metastable state and then eventually decays to 
an equilibrium state. How long does the system remain in the metastable 
state? How does this state decay to the equilibrium state? What are 
the reversal mechanisms responsible for such a decay? All these questions are 
nicely explained by the classical nucleation and droplet theory, according to which in a spin-$1/2$ Ising system below $T_{\rm c}$, small clusters or droplets of down spins are dispersed in the sea of 
up spins~\cite{gunton}. The number of such droplets with size $l$ 
($l$ denotes the number of spins ``$-1$'') is given by the Boltzmann distribution
\begin{equation}\label{eq:1}
n_l = Ne^{-\beta E_l},
\end{equation}
where $N$ is a normalization factor, $\beta=1/k_{\rm B} T$, $k_{\rm B}$ is the Boltzmann constant, $T$ the temperature, and $E_{l}$ the free energy of a droplet formation of size $l$. In fact, $E_{l}$ comes from the contribution of bulk and surface energies
\begin{equation}\label{eq:2}
	E_l=-2hl+\sigma l^{(d-1)/d},
\end{equation}
where $d$ is the space dimension, $-2hl$ corresponds to the bulk energy required to flip $l$ number of spins in a magnetic field $h$ and the second term expresses the energy associated with the surface tension $\sigma$ of the droplet. Historically, in classical nucleation theory Allen-Cahn considered the Langevin equation of the droplet dynamics which obviously 
involves the free energy. This is nicely reviewed in Ref.~\cite{gunton}

Classical nucleation theory~\cite{becker,gunton} assumes that the droplets are mostly spherical-like in the $d$-dimensional space. Thus, 
the radius of a droplet size $l$ will be proportional to $\sim l^{1/d}$. 
The surface area will then be proportional to $\sim l^{(d-1)/d}$. Due to the opposite polarity of the two terms in Eq.~(\ref{eq:2}) there is a competition between the two energies. As a consequence there must be a critical size of the droplet $l_{\rm c}$ (with a radius $R_{\rm c}$) for which the free energy is maximum, as given below
\begin{equation}\label{eq:3}
l_{\rm c}=\Bigg(\frac{\sigma(d-1)}{2d|h|}\Bigg)^d.
\end{equation} 
In the vicinity of the free-energy maximum
interesting dynamical phenomena are observed. Only the supercritical droplets 
having size $l > l_{\rm c}$ will be energetically favored to grow (reducing the free energy) and eventually 
engulf the whole system helping the reversal of magnetization. In contrast, the
subcritical droplets ($l < l_{\rm c}$) are found to shrink reducing the free 
energy and they are unable to take part in the reversal process. From this explanation 
of the droplet theory it is obvious why the system consumes a certain period of time to leave the metastable state. 

So the maximum energy associated with the critical size of 
the droplets is
\begin{equation}\label{eq:4}
E_l \Big |_{l=l_{\rm c}}=E_{\rm c}=\frac{K_d\sigma^d}{h^{d-1}},
\end{equation}
where $K_d= \frac{1}{d} \Big( \frac{d-1}{2d} \Big)^{d-1}$ is a $d$-dependent constant term. Now what is the growth rate of such 
droplets for different strengths of the applied field? This problem was analyzed 
by Becker and D\"{o}ring~\cite{becker,gunton}. The basic assumption of their 
analysis is that the time evolution of the number of droplets is 
only due to an evaporation-condensation mechanism in which a droplet 
either loses or gains a single spin without any type of coagulation or 
other interactions. The main result is that the number 
of droplets formed per unit time and volume (nucleation rate), $I$, is 
largely dependent on $E_{\rm c}$
\begin{equation}\label{eq:5}
I=I_{0} e^{-E_{\rm c}/k_{\rm B}T},
\end{equation}
where $I_{0}$ is the rate prefactor. Thus, the metastable lifetime in the nucleation 
regime, $\tau_{\rm (nr)}$, can easily be obtained from the inverse 
nucleation rate $I$ via
\begin{equation}\label{eq:6}
	\tau_{\rm (nr)} \sim I^{-1} \sim \exp\Bigg(\frac{K_{d}\sigma^d}
	{k_{\rm B}Th^{d-1}}\Bigg).
	\end{equation}

As the applied-field strength is further increased, a different scenario for the reversal mechanism is observed: Many critical droplets grow simultaneously ($l^{1/d} \sim t$) and coalesce with each other resulting in complete reversal of spins. Clearly the rate of change in the magnetization is proportional to the 
droplet-size variation
\begin{equation}\label{eq:7}
\frac{dm}{dt} \sim \frac{dl}{dt} \sim It^d \rightarrow \Delta m \sim I  \frac{\tau_{(cr)}^{d+1}}{d+1},
\end{equation}
where $\tau_{\rm (cr)}$ is the metastable lifetime at the coalescence
regime. So for a fixed change in the magnetization $\Delta m$,  the metastable 
lifetime in the coalescence regime is
\begin{equation}\label{eq:8}
	\tau_{\rm (cr)} \sim I^{-\frac{1}{d+1}} \sim \exp\Bigg(\frac{K_d\sigma^d}
	{k_{\rm B}T(d+1)h^{d-1}}\Bigg).
	\end{equation}

From the above description and in particular from Eqs.~(\ref{eq:6}) and (\ref{eq:8}) we can securely deduce that the metastable lifetime logarithm is described by a fair straight line when depicted as a function of the inverse applied-field strength. In addition to that, the slope of this line is expected to be smaller in the coalescence regime when compared to that in nucleation. 

\section{Model and Simulation Details}
\label{model}

The spin-$s$ Blume-Capel model~\cite{blume}, where $s$ assumes integer or half-integer values of the spin, is described by the Hamiltonian 
\begin{equation}\label{eq:9}
\mathcal{H} = - \frac{1}{s^2}\;J \sum_{\langle i,j\rangle} s_i^z s_j^z + \frac{1}{s^2}\; \Delta \sum_i (s_i^z)^2 - \frac{1}{s}\; h \sum_i s_i^z,
\end{equation}
where $s_i^z$ denotes the $z$-component of the spin which 
can take values from $-s$ to $+s$ through unit steps. For example, this means that
for the $s = 5/2$ system, the normalized spin components $s_i^z = \{1,3/5,1/5,-1/5,-3/5,-1\}$ have been considered, instead of the usual $s_i^z  = \{5/2,3/2,1/2,-1/2,-3/2,-5/2\}$. Obviously, for $s=1/2$ and $\Delta=0$ the original Ising model's Hamiltonian is recovered. Seven spin systems have been studied in the present work, namely the systems with $s = 1/2,1,3/2,2,5/2,3$, and $7/2$. Returning to the description of Hamiltonian~(\ref{eq:9}), we note the following: (i) The first term is the energy interaction between nearest-neighbor spins with uniform ferromagnetic coupling ($J > 0$). (ii) The second term models the single-ion anisotropy (or crystal-field coupling) $\Delta$ and controls the density of vacancies ($s_i^z = 0$) for the models with $s=1,2$, and $3$. (iii) The third term is the well-known Zeeman energy describing the interaction of an externally applied uniform magnetic field $h$ along the $z$-direction with each individual spin. Both $\Delta$ and $h$ has been measured in units of $J$ which sets the energy scale, and as usual we have set $J = k_{\rm B} = 1$ to fix the temperature scale.  Extensive results regarding the critical properties of the general spin$>1$ Blume-Capel model can be found in the works by Plascak and collaborators~\cite{plascak93,plascak03}. For the particular spin-$1$ case we refer the reader to Refs.~\cite{fytas17,fytas18,fytas20}.

All the numerical data shown in the following Section were obtained via Monte Carlo simulations on a two-dimensional $L \times L$ square lattice, where $L = 100$, with periodic boundary conditions. As a side remark we note here that the first and most extensive numerical work on the problem is due to Acharyya and Stauffer~\cite{stauffer} were sizes up to $L = 2048$ were simulated on the square-lattice Ising ferromagnet. In that work a special technique, the so called multispin coding, was applied, where each bit of computer memory is filled by either $1$ (for spins up) and $0$ (for spins down). Moreover a geometric parallelization was employed to facilitate the simulations, meaning that the whole lattice was divided into several strips and each strip was updated in different nodes of a CRAY-T3E supercomputer. However, in the present study of the general spin-$s$ models and in contrast to the spin-$1/2$ case, spin projections are not restricted to special values so that one cannot directly apply the method of multispin coding. In this respect, we consider the moderate size $L = 100$ as a fair compromise of the available computational facilities that will allow for a safe estimation of the properties under study, see also Ref.~\cite{fytas10}. Of course, more extensive studies including larger system sized would be welcome but we believe that the main conclusions of our work will remain solid.

For the Blume-Capel models we have fixed the single-ion anisotropy to $\Delta = 0.5$ throughout this work. Note that the location of tricritical points in the spin-$1$ and $3/2$ models lie close to $\Delta_{\rm t} \approx 1.966$ and $1.986$, respectively~\cite{fytas17,plascak03}. Thus, we expect that $\Delta = 0.5$ is a reasonable choice for the current needs, since it guarantees that we stay well within the second-order transition regime of the phase diagram. 

In our numerical protocol the system is initially considered to be in the perfectly ordered state where all $s_{i}^{z} = +1$. Then a random updating scheme using the Metropolis single spin-flip algorithm is put into play which follows the discrete time Markov chain based on the principle of detailed balance~\cite{metrop,binderbook}. Let us briefly summarize the Metropolis algorithm's steps below:  

\begin{enumerate}
	
\item A lattice site is selected randomly among the $L \times L$ options.
	
\item The spin variable located at the selected site is flipped, keeping the other spins in the system fixed. The updated spin state may then be in any of the $2s+1$ possible states between $-s$ to $+s$ which have been determined randomly with equal probability using the uniformly distributed random numbers between $0$ and $1$.
	
\item The energy change originating from this spin-flip operation is calculated using the Hamiltonian of Eq.~(\ref{eq:9}) as follows: $\Delta \mathcal{H}=\mathcal{H}_{\rm a}-\mathcal{H}_{\rm o}$, 
where $\mathcal{H}_{\rm a}$ denotes the system's energy after the trial switch of the selected spin and $\mathcal{H}_{\rm o}$ corresponds to the total energy of the old spin-configuration system. The probability to accept the proposed spin update is given by
\begin{equation}
\label{eq:10}
W_{M}\left(s_{i}^{z}\rightarrow s_{i}^{z'} \right)=
\begin{cases}
\exp(-\Delta \mathcal{H}/k_{\rm B}T),       & \text{if } \mathcal{H}_{a} \geq \mathcal{H}_{o}, \\
1,        & \text{if } \mathcal{H}_{\rm a} < \mathcal{H}_{\rm o}.
\end{cases}
\end{equation}
	
\item If the energy is lowered, the newly chosen value of the spin projection is always accepted. 
	
\item If the energy is increased, a random number $R$ is generated, such that $0 < R < 1$: If $R$ is less than or equal to the calculated Metropolis transition probability the selected spin is flipped. Otherwise, the old spin configuration remains unchanged. 
	
\end{enumerate}

This process corresponds to an update of a single spin. One Monte Carlo step per spin (MCSS) consists of a $10^{4}$ number of such updates which act as the unit of time throughout this study. After the completion of each MCSS, the magnetization is determined via
\begin{equation}\label{eq:11}
	m(t)= \frac{1}{N} \sum_{i=1}^{N} s_i^z,
\end{equation}
where $t$ is the time in units of MCSS and $N=L^{2}$ the number of lattice points. The averaged magnetization is then calculated by $m(t)$ over the last $2 \times 10^{4}$ MCSS, where the first $5\times 10^3$ steps are discarded during the thermalization process. 

For the determination of suitable pseudocritical points the lattice was allowed to evolve for $2.5 \times 10^{4} $ MCSS in the absence of any externally applied magnetic field. The pseudocritical temperatures were estimated from the peak location of the finite-lattice magnetic susceptibility curves $\chi$, computed by the response function of the magnetization via
$\chi = \beta N[\langle m^{2}\rangle-\langle m\rangle^{2}]$. These estimates were found to be in good agreement with previous literature~\cite{fytas10}, as also discussed below.

Closing this Section, we would like to make a few comments on the numerical approach and error estimation. Although the Metropolis algorithm may not
be the optimum choice for studying critical properties of Ising and Blume-Capel models with higher spin-values~\cite{bekhechi97,plascak02}, in the present work it is a safe and convenient vehicle, as we are only interested in the metastable behavior well below the critical point. Additionally, we expect the main conclusions of our work to be qualitatively insensitive to the use of other algorithms. At a quantitative level however, the relaxation (or metastable) time can change depending on the chosen algorithm. For example, it is well-known that local- and cluster-update algorithms belong to different dynamic universality classes~\cite{newmanbook} and distinct values of metastable times should be expected. Finally, with respect to error estimation we have used  the standard simple block averaging method~\cite{binderbook}. Note that in some cases these error bars are not visible due to being of the order of symbol sizes.

\section{Numerical Results}
\label{results}

The main aim of this paper is to understand the behavior of metastable states in
spin-$s$ Ising and Blume-Capel models. According to the Becker-D\"{o}ring 
theory, the metastable state of the spin-$1/2$ Ising 
ferromagnet decays through the three different reversal mechanisms of 
the spin: strong-field regime, coalescence regime, and nucleation regime). 
These scenarios have been outlined above in Sec.~\ref{theory}.
Before presenting our main results we would like to remind the reader that we mainly focus on the coalescence and nucleation regimes and attempt to provide some clear answers to the following open questions: Does the Becker-D\"{o}ring analysis hold for the general case of the spin-$s$ Ising model? What are the effects of single-ion anisotropy? 

\begin{figure}[h!]
\subfigure[\hspace{0 cm}] {\includegraphics[width=8cm]{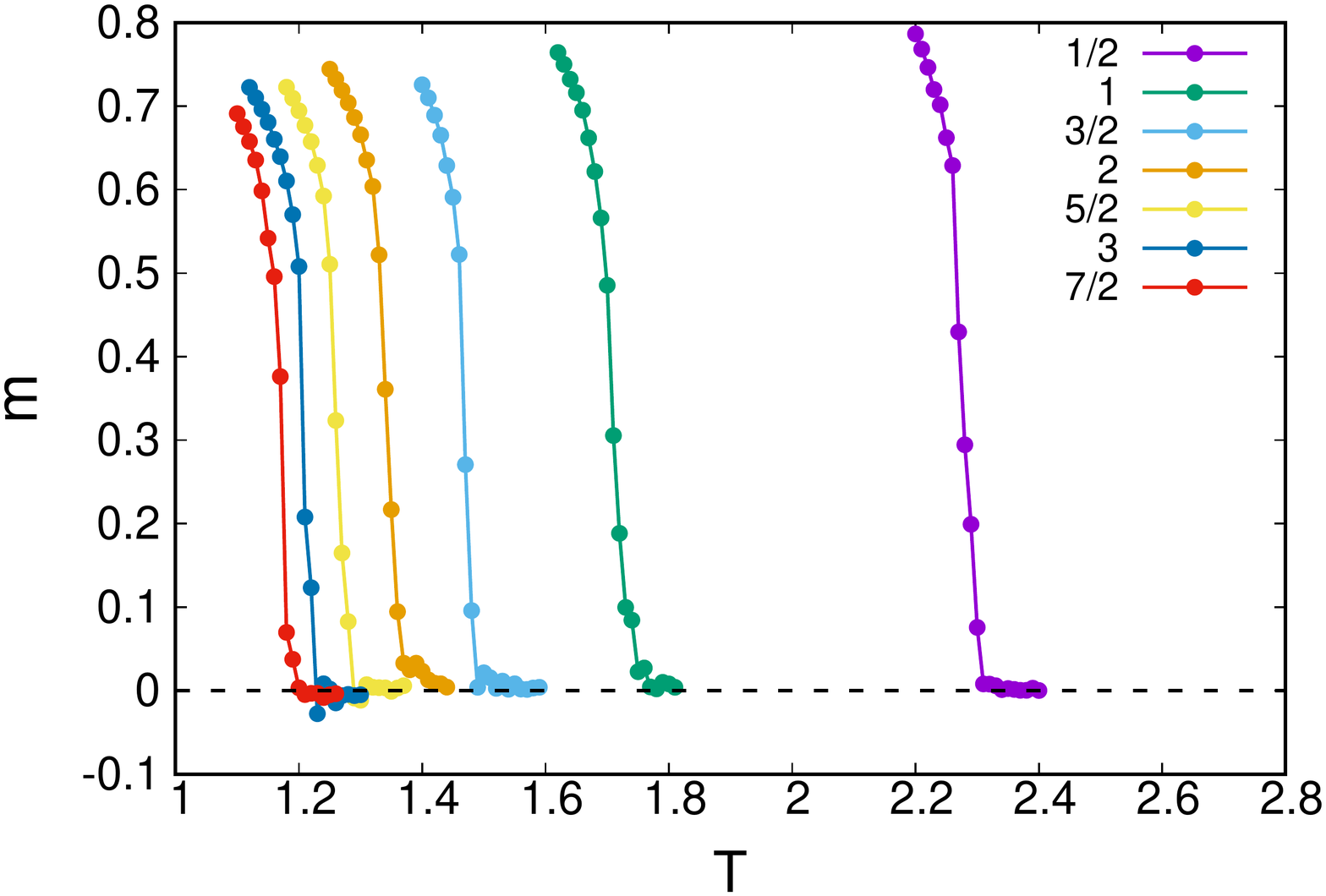}}
\subfigure[\hspace{0 cm}] {\includegraphics[width=8cm]{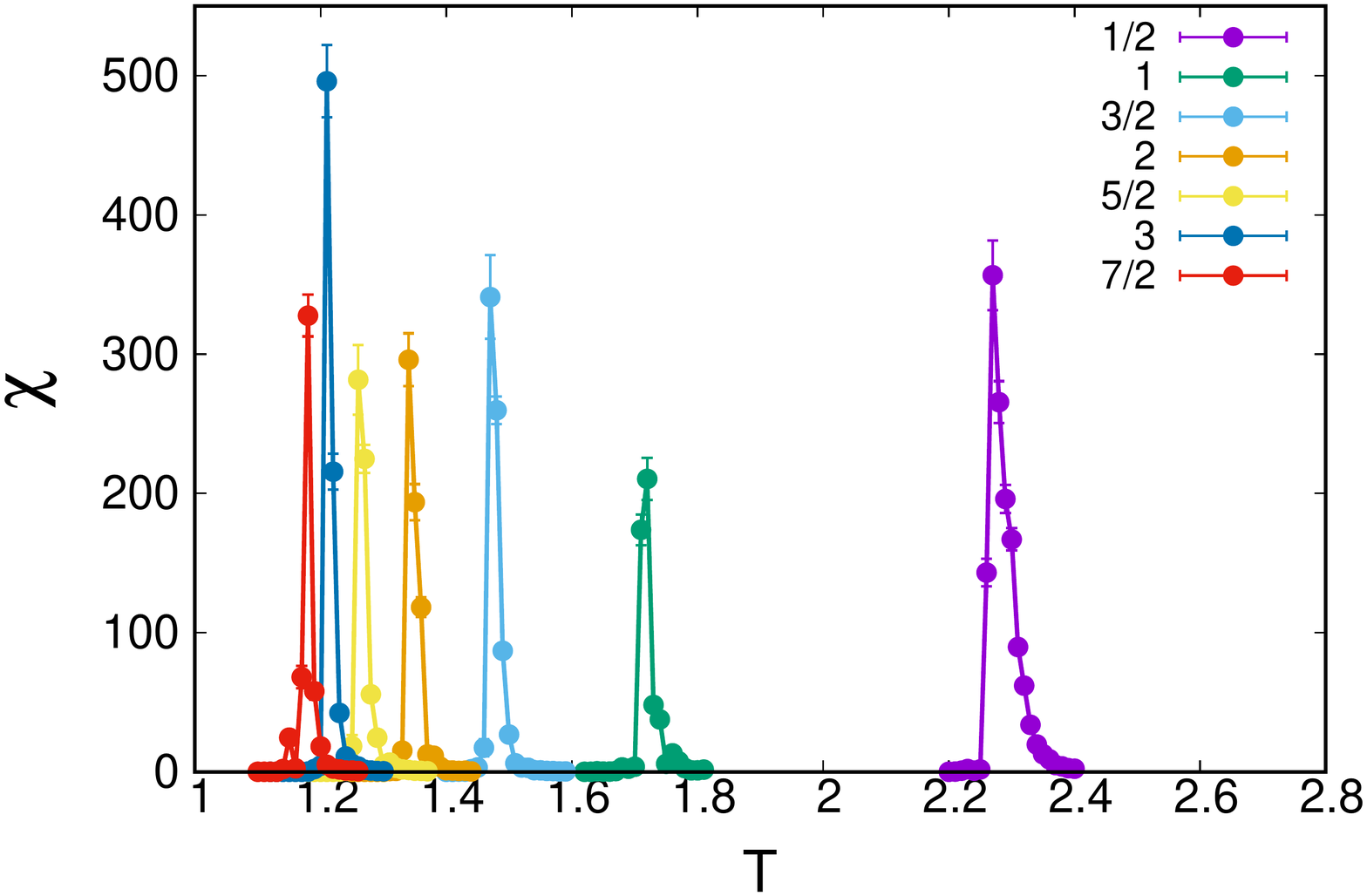}}\\
\caption{Temperature dependence of the magnetization (a) and magnetic susceptibility (b) for the spin-$s$ Ising models, where $ s= 1/2, 1, 3/2, 2, 5/2,3$, and $7/2$. Results averaged over $200$ samples. In some of the data points error bars are not visible as they are smaller than the symbol size.}\label{istc}
\end{figure}

\begin{figure}[h!]
\subfigure[\hspace{0 cm}] {\includegraphics[width=8cm]{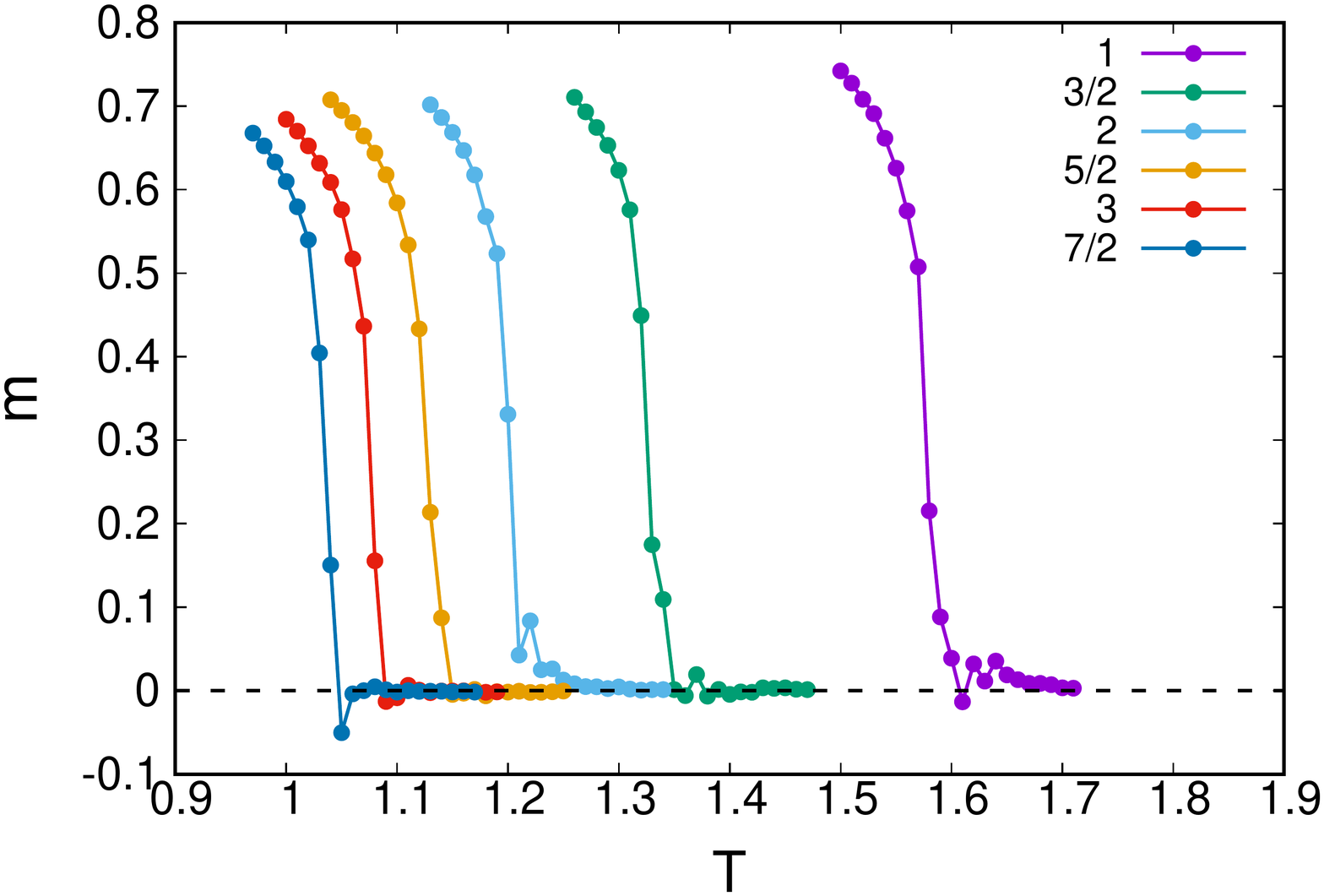}}
\subfigure[\hspace{0 cm}] {\includegraphics[width=8cm]{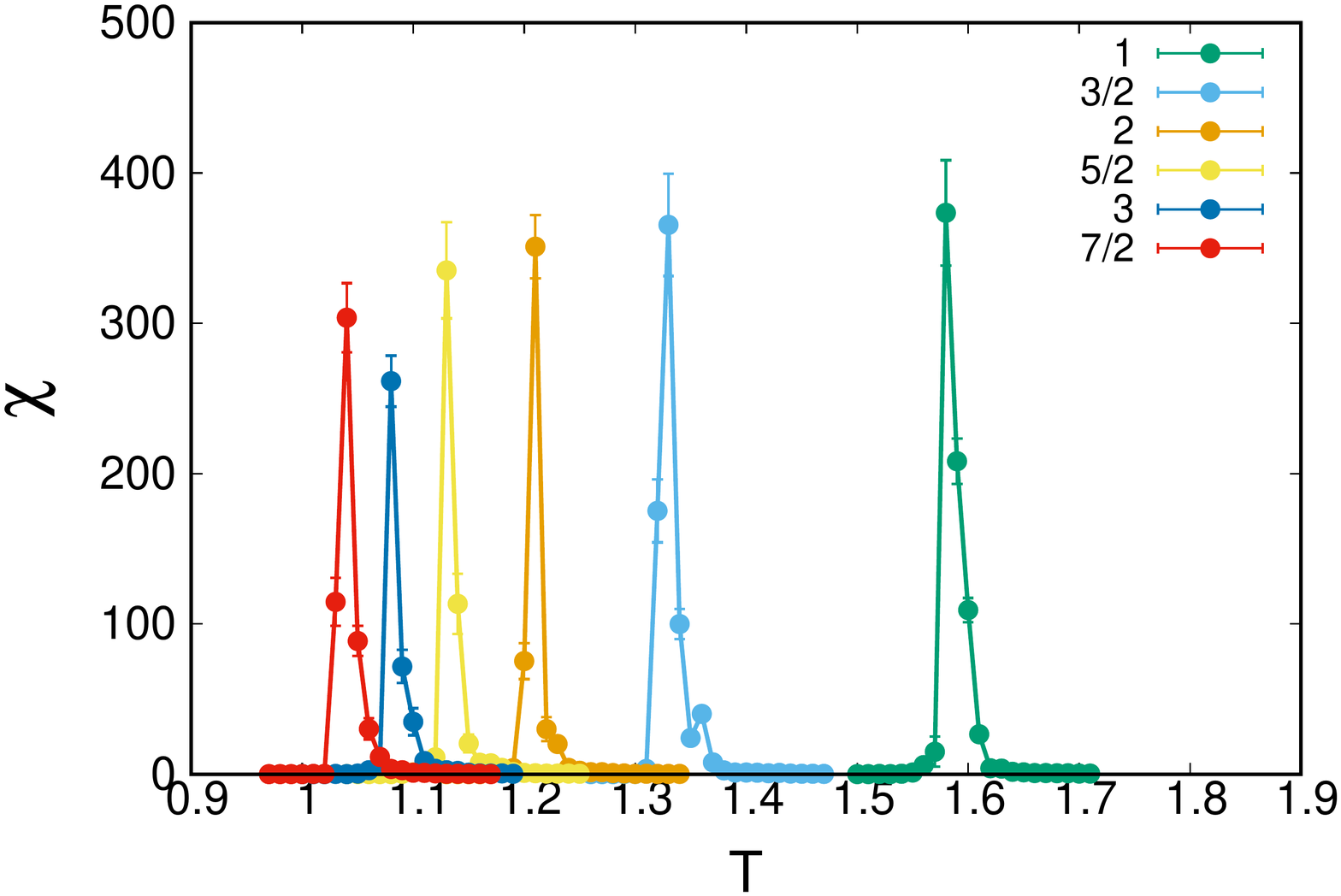}}\\
\caption{The same as in Fig.~\ref{istc} for the spin-$s$ Blume-Capel models. Again, in some of the data points error bars are not discernible due to being smaller than the symbol size.}\label{bctc}
\end{figure}

Ideally, we would like to have a rough estimate of the critical 
temperature $T_{\rm c}$ of the normalized spin-$s$ Ising and Blume-Capel models. The simplest way is to perform a finite-size scaling study based on the shift behavior of some suitably defined pseudocritical temperatures, $T_{L}^{\ast}$, \emph{i.e.}, those corresponding to susceptibility peaks~\cite{binderbook,ferrenberg}. However, the situation is much easier in the sense that we do not really need to know exactly $T_{\rm c}$ for each model. To explore the Becker-D\"{o}ring theory, the system must simply be kept below the critical temperature, say around $\sim 0.7T_{\rm c} $. Thus, if we simply consider the pseudocritical temperature of the susceptibility for a moderate system, this should be enough to ensure that the ratio $0.7T_{L}^{\ast} < T_{\rm c}$. In this respect we studied the variation of the equilibrium magnetization $m$ and the corresponding susceptibility $\chi$ for both Ising and Blume-Capel systems, as shown in Figs.-\ref{istc} and \ref{bctc}, respectively. 

A few comments are in order: (i) The susceptibility was checked 
by varying the temperature in steps of $\delta T = 0.01$ so that the maximum 
error associated with the approximate $T_L^{\ast}$ is of the order $\sim 10^{-2}$. In Tab.~\ref{tctable}  a summary of approximate pseudocritical temperatures (considered up to second decimal place) is provided for the spin-$s$ Ising and Blume-Capel models. For a direct comparison note that a value $T_{\rm c}(\Delta=0.5) = 1.564(3)$ was obtained in Ref.~\cite{fytas10} for the spin-$1$ Blume-Capel model. (ii) From Fig.~\ref{istc} we observe that the critical temperature decreases with increasing $s$. In the spin-$1/2$ Ising system, the spin can 
either access the state ``$+1$'' or ``$-1$''. As the number of spin components 
increases the system walks through some intermediate accessible 
states between ``$+1$'' and``$-1$''. Then the activation energy needed 
to flip the spin from ``$+1$'' or ``$-1$'' via some intermediate state will be 
much smaller compared to the direct flipping. Furthermore, 
in the disordered state, the spin-$s$ system will be equally distributed among all 
of its accessible states. For these reasons the system consumes 
lower energy (thermal activation-energy) to be driven into a fully 
disordered state. (iii) For the case of the Blume-Capel model, $T_L^{\ast}$ is found to decrease in the presence of anisotropy 
compared to the spin-$s$ Ising model. The presence of 
anisotropy favors energetically some of the intermediate states. Say for example for the spin-$5/2$ Blume-Capel model, 
normalised spin states ``$\pm 1/5$'' will be more favored 
than ``$\pm 3/5$'' which are again more favorable than ``$\pm 1$'' -- ]note that this statement is true for positive values of $\Delta$, according to the description of Hamiltonian~(\ref{eq:1}). 
So the thermal energy required to make an anisotropic system disordered 
is smaller compared to the one needed for an isotropic Ising system.

\begin{figure}[h!]
\subfigure[\hspace{0 cm}] {\includegraphics[width=8cm]{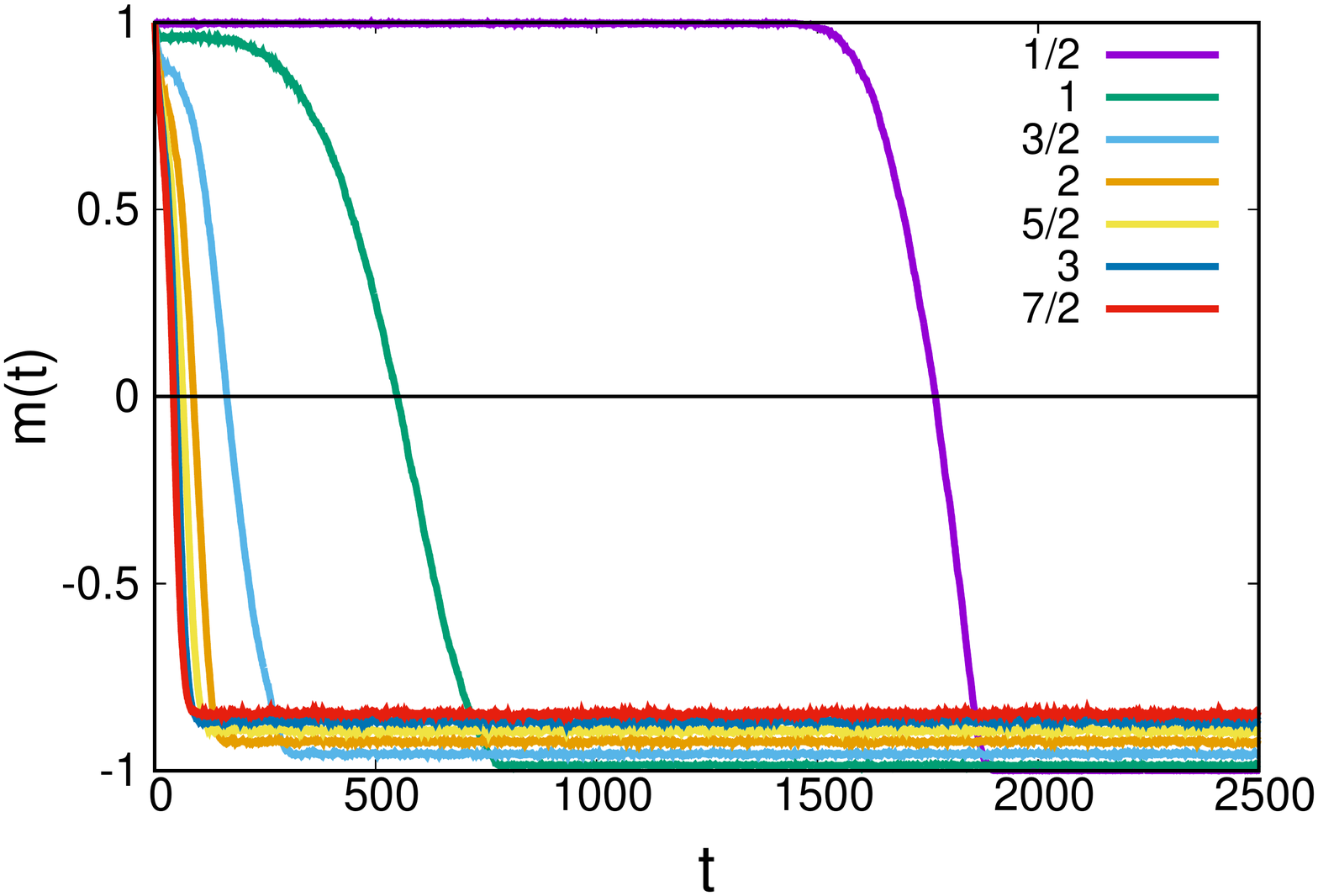}}
\subfigure[\hspace{0 cm}] {\includegraphics[width=8cm]{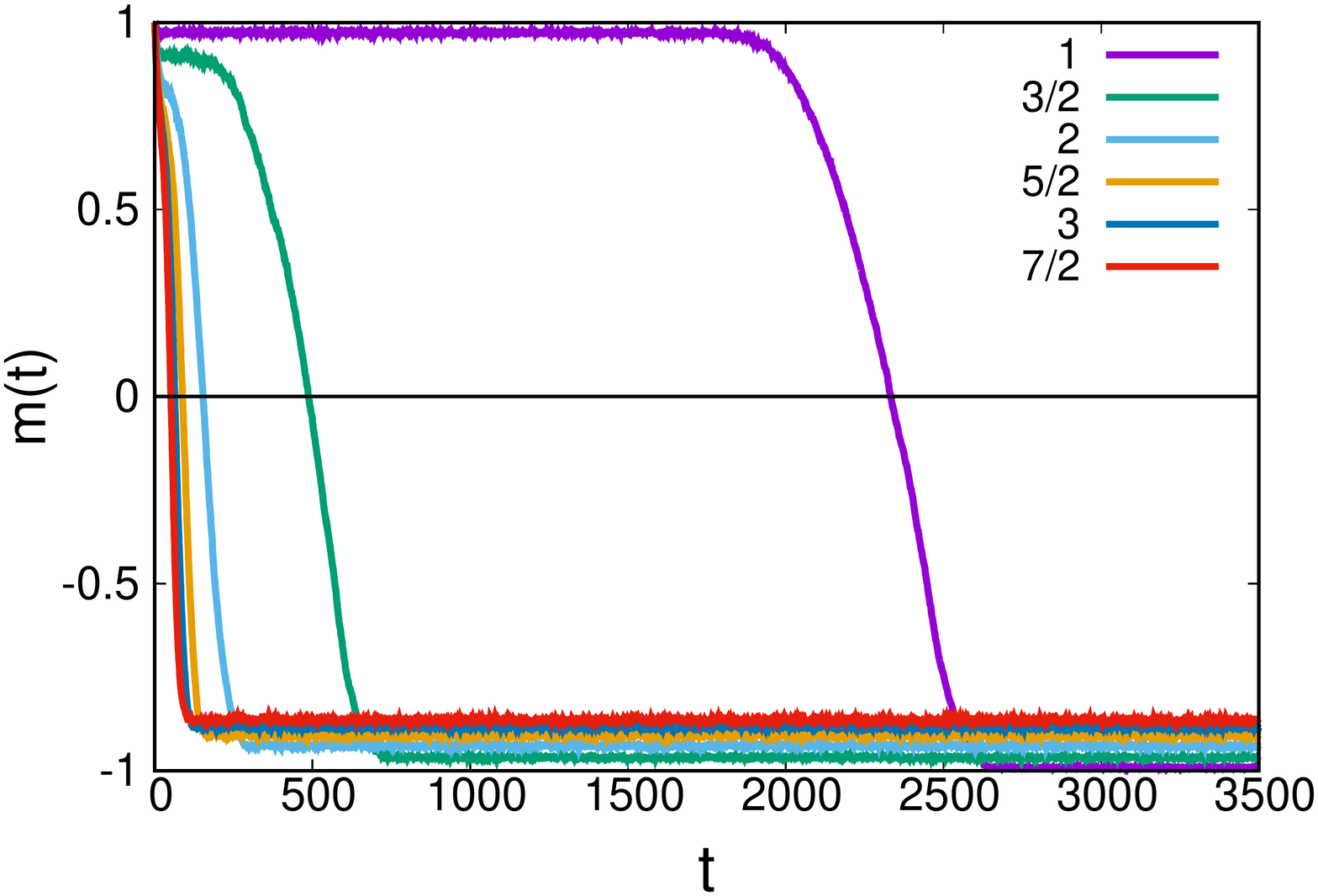}}\\
\caption{(a) Variation of the magnetization with time (in MCSS) for Ising models at $T=1.0$ and $h = -0.4$. Reversal times for the spin-$\{1/2, 1, 3/2, 2, 5/2,3, 7/2\}$ systems are $\tau= 1767, 549, 163, 87, 63, 48$, and $42$ MCSS, respectively. (b) Analogous results for Blume-Capel models at $T=0.8$ and $h = -0.4$. Reversal times for the spin-$\{1,3/2,2,5/2,3,7/2\}$ systems are $\tau= 2332, 487, 153, 88, 63$, and $51 $ MCSS, respectively. In both panels typical benchmark curves obtained from a single run are shown for illustrative reasons.}\label{magtime}
\end{figure}

Below the critical temperature and if a very small magnetic field (in opposite 
direction to the spin moment) is applied to a ferromagnet, the system 
will enter into a metastable state with some positive magnetization. Then,  
after a certain period of time it will reach an equilibrium state through the 
decay of this metastable state. The time needed by the system to escape from 
the metastable state is known as the metastable lifetime which is 
also referred here as the reversal time $\tau$ of the magnetization. 
Quantitatively $\tau$ has been defined here as the time at which 
the magnetization changes sign ($m(t)\sim 0$). In Fig.~\ref{magtime}(a), 
we have checked the metastable lifetime of the spin-$s$ Ising system 
in the presence of a small negative magnetic field $h=-0.4$ and at $T= 1.0$ which is well below the critical temperature of the
spin-$7/2$ case, so that $T < T_{\rm c}$ for all systems. The main observation here is that at a particular temperature 
and in the presence of a uniform magnetic field the reversal time is found to 
decrease with the increasing $s$. This is due to the increase in the thermal fluctuations which is connected to the fact that as we go to 
the higher spin values $T_{\rm c}$ decreases and gradually approaches the fixed 
temperature $T=1.0$. Similar results are presented also for the corresponding Blume-Capel models in Fig.~\ref{magtime}(b) for the same magnetic field value and a temperature $T=0.8$ as outlined in the figure.

\begin{figure}[h!]
\subfigure[$\; s=1/2$, $T = 1.59$. \hspace{0 cm}] {\includegraphics[width=8cm]{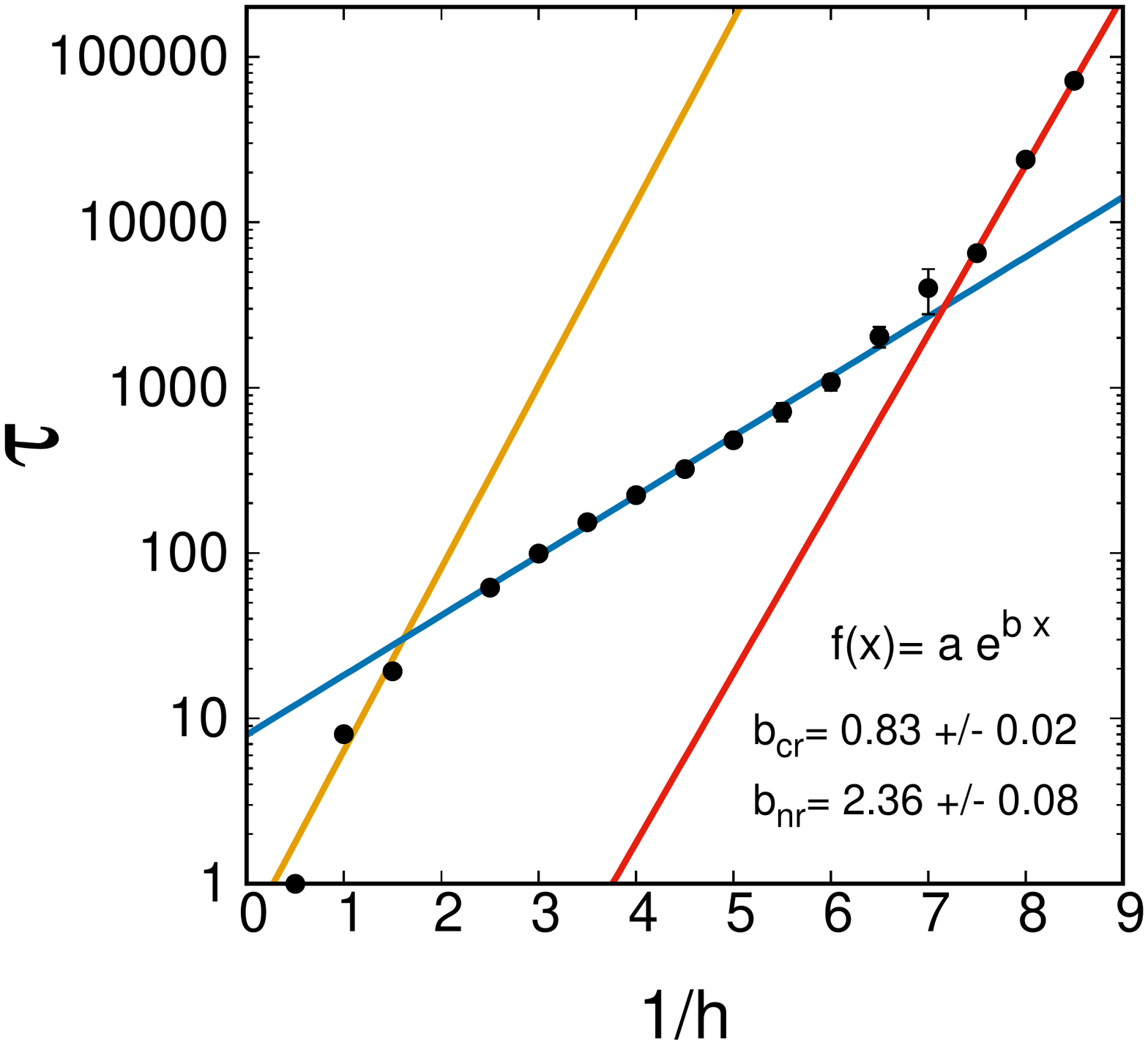}}
\subfigure[$\; s = 2$, $T = 0.94$.\hspace{0 cm}] {\includegraphics[width=8cm]{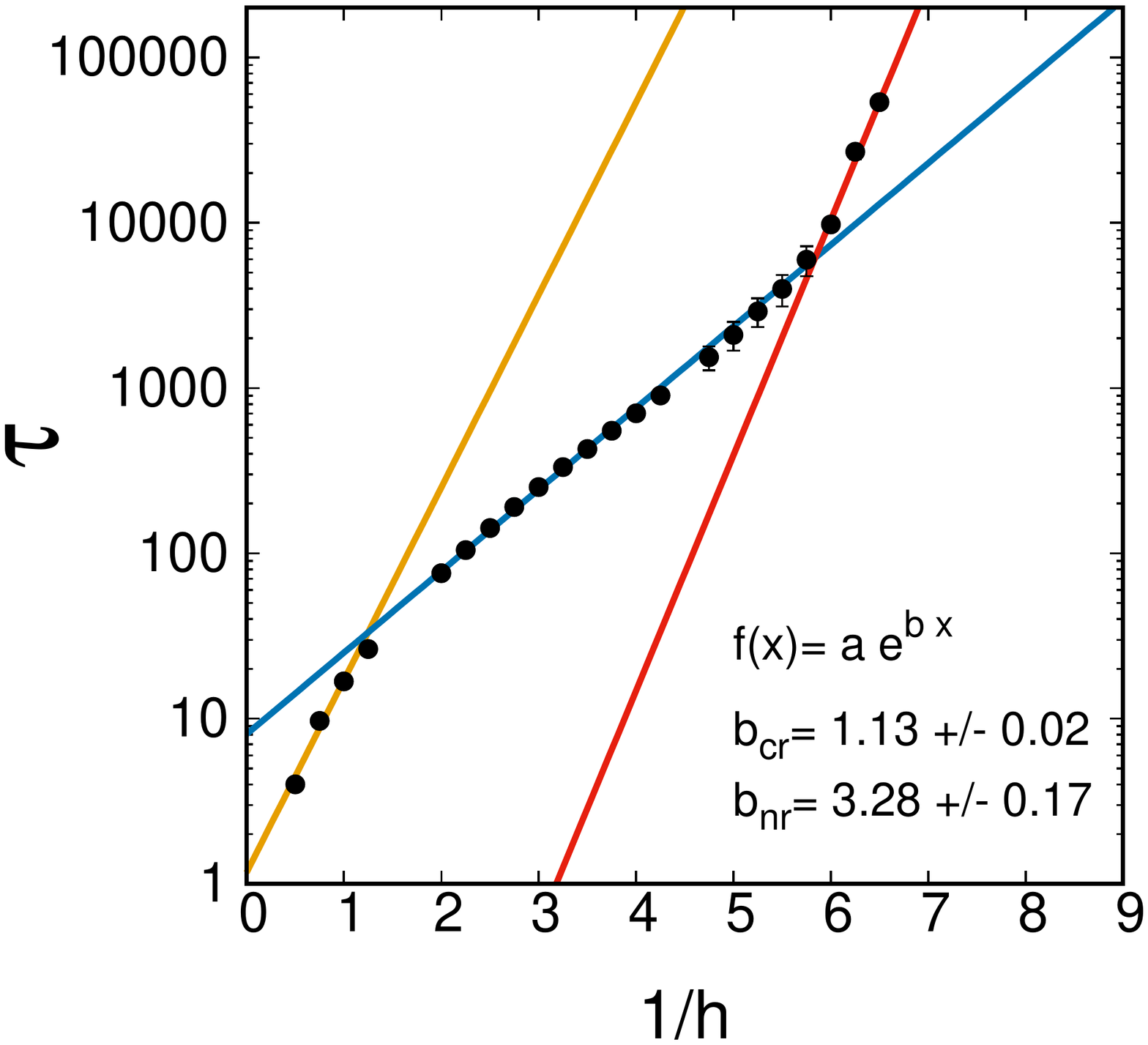}}\\
\subfigure[$\; s = 5/2$, $T = 0.88$.\hspace{0 cm}] {\includegraphics[width=8cm]{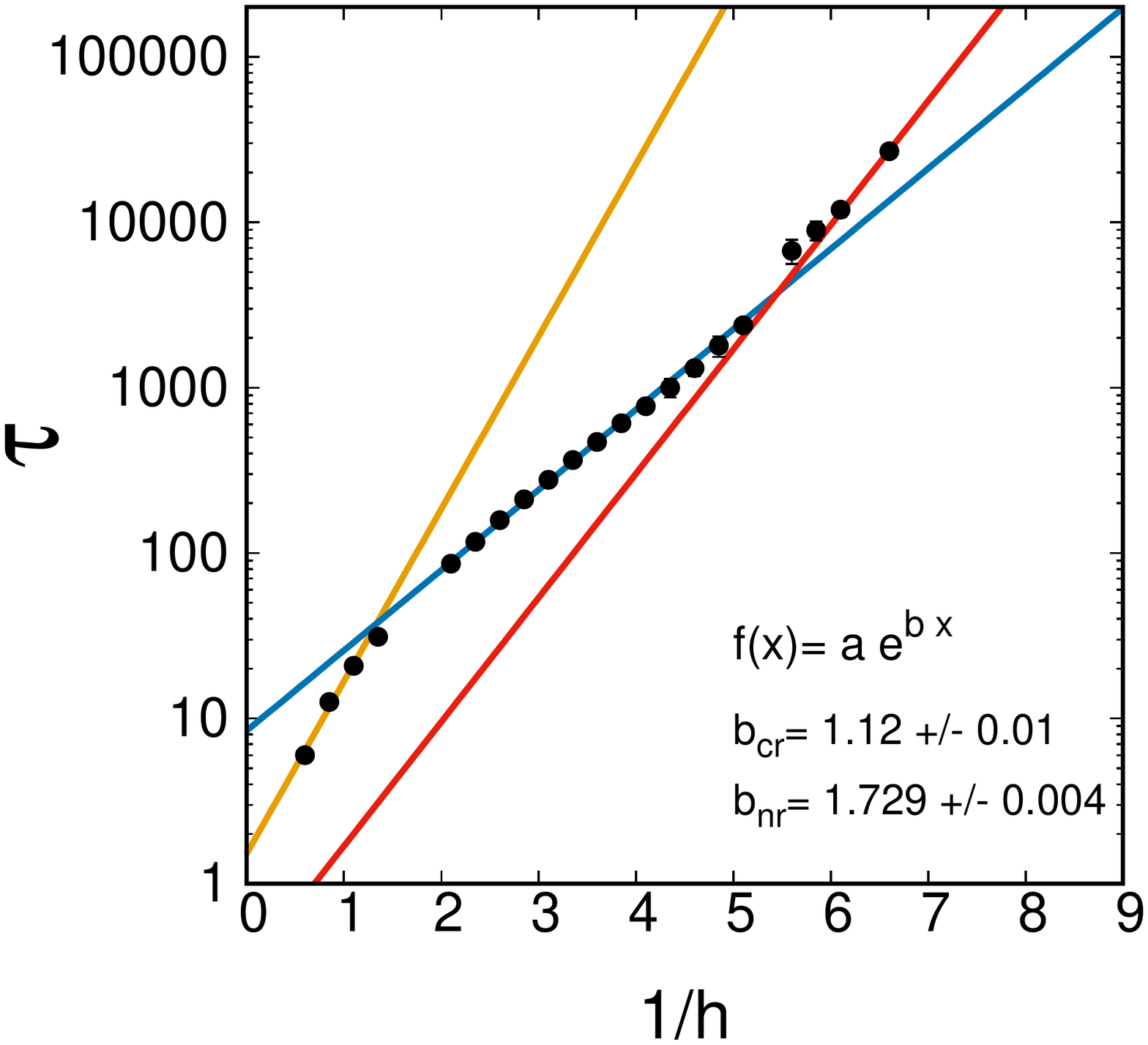}}
\subfigure[$\; s = 3$, $T = 0.85$.\hspace{0 cm}] {\includegraphics[width=8cm]{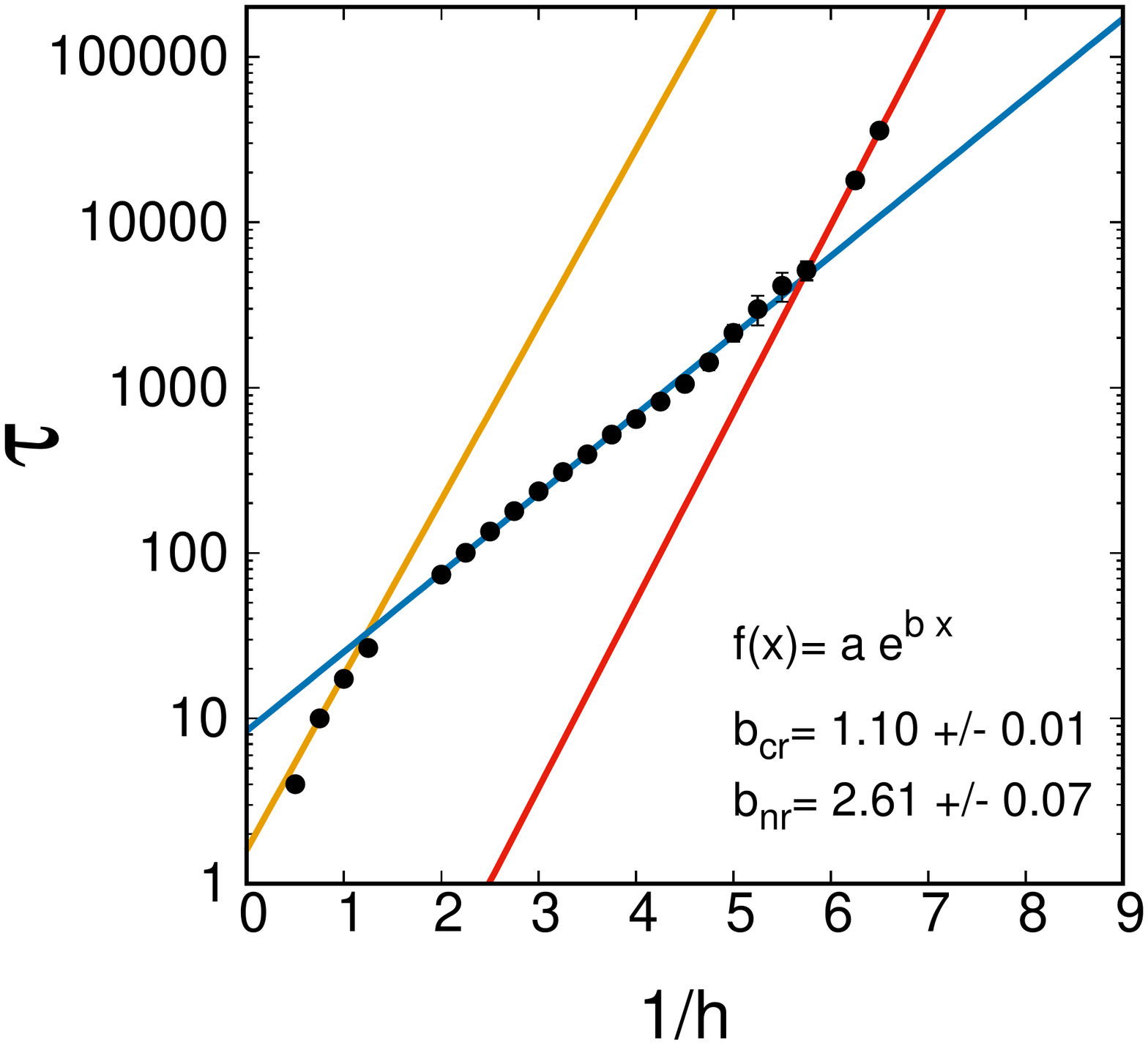}} 
 
\caption{Mean reversal time as a function of the inverse magnetic field at $T= 0.7T_{L}^{\ast}$ for four spin-$s$ Ising models, as indicated in the panels. Results averaged over $1000$ samples. In cases where the error bars are not visible, this is due to being smaller than the symbol size used. Note the appearance of three different regimes: (i) strong-field regime (yellow line), (ii) coalescence regime (blue line), and (iii) nucleation regime (red line), which are identified with different slopes. Note the logarithmic scale in the vertical axis. See also Tab.~\ref{chisquare-is}.}\label{isingbeck}
\end{figure}

Fixing now the temperature to $T = 0.7T_{L}^{\ast}$, we present in Fig.~\ref{isingbeck} a variation of the mean 
reversal time as a function of the inverse magnetic field for the spin-$1/2$, $2$, $5/2$, and $3$ Ising models. Three different regimes with distinct slopes are clearly identified (see also discussion in the figure panels). For the fits shown we have implemented the standard $\chi^2$ test for goodness
of fit. Specifically, the $p$ value of our $\chi^2$ test, also known
as $Q$, see \emph{e.g.}, Ref.~\cite{press}, is the probability of finding a
$\chi^2$ value which is even larger than the one actually found
from our data. Recall that this probability is computed by
assuming (1) Gaussian statistics and (2) the correctness of
the fit’s functional form. We consider a fit as being fair only
if $10\% < Q < 90\%$. In Tab.~\ref{chisquare-is} the results with respect to the quality of our fits are given and provide a strong credibility test.

Qualitatively one can clearly argue that the Becker-D\"{o}ring analysis 
holds for the general spin-$s$ Ising systems studied. An interesting point is that, although each system is kept at $T= 0.7T_{L}^{\ast}$, the metastable lifetime 
rises for $s > 1/2$ (see Fig.~\ref{isingbeck}(b) - (d)) with 
respect to that for $s=1/2$ (see Fig.~\ref{isingbeck}(a)). As we discussed previously, systems with $s > 1/2$ have some intermediate states 
between ``$+1$'' and ``$-1$''. Now if we go back to the droplet theory, the bulk 
energy term in Eq.~(\ref{eq:2}) will be modified for the spin-$s > 1/2$ case, since the energy needed to flip the droplet of some intermediate state will be less than the one needed to flip the droplet of spin ``$+1$''. The droplet's formation-energy will then be $E_l=-ahl+\sigma l^{(d-1)/d}$, where $0 < a \le 2$. Obviously, $a=2$ corresponds to the spin-$1/2$ Ising system. Thus, the critical 
droplet size will be $l_{\rm c}=[\sigma(d-1)/(a d|h|)]^d$. 
Clearly, the critical size $l_{\rm c}$ becomes larger for systems with $s > 1/2$ resulting in longer times needed for the system to escape the metastable state. Also, the $d$-dependent constant term $K_{\rm d}$ increases since the denominator factor $2$ in Eq.~(\ref{eq:3})
will be replaced by $a$. Finally, from Eqs.~(\ref{eq:6}) and (\ref{eq:8}) it can be confirmed also that the reversal time increases.
 
\begin{figure}[h!]
\subfigure[$\; s = 3/2$, $T= 0.93$. \hspace{0 cm}] {\includegraphics[width=8cm]{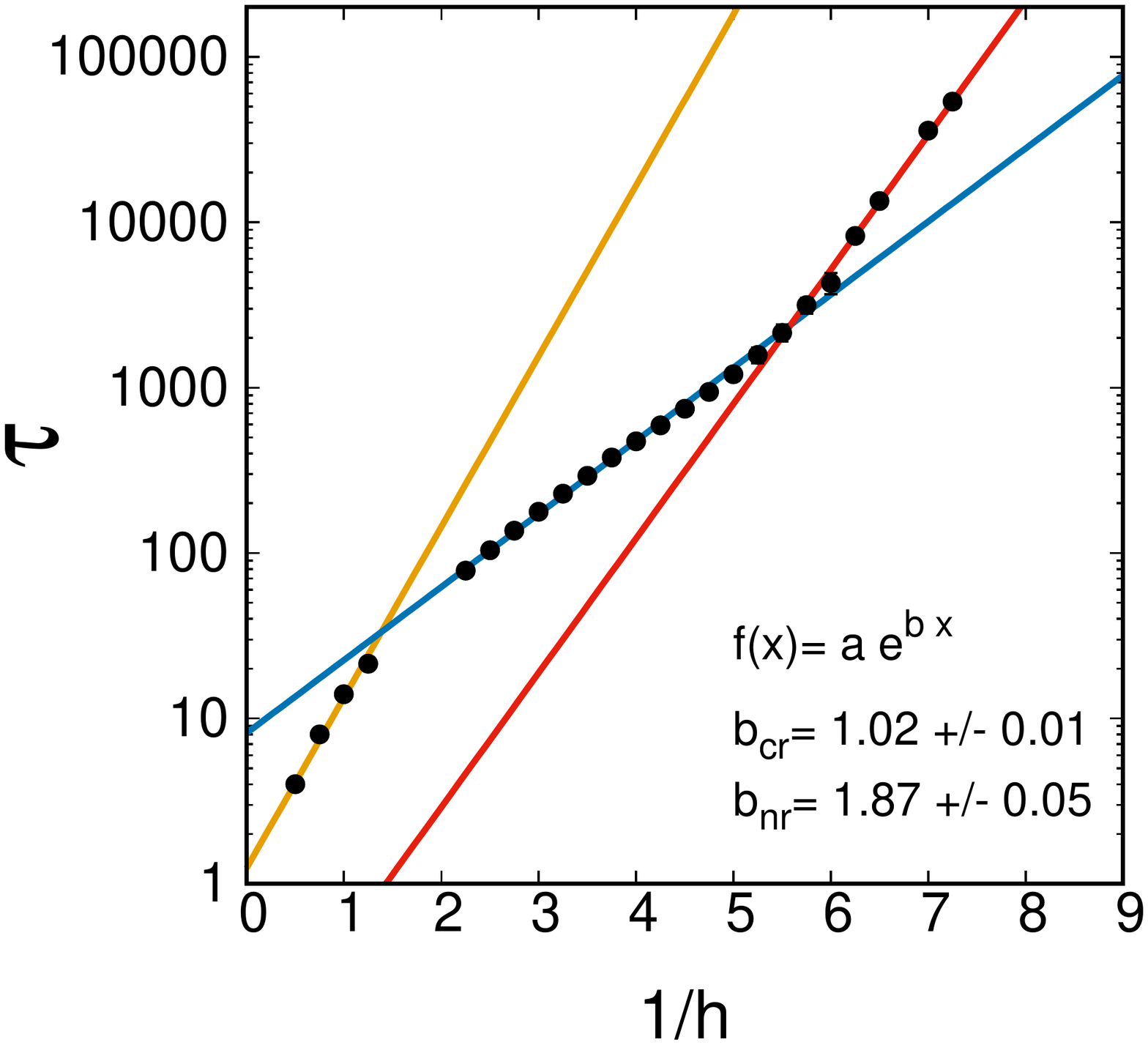}}
\subfigure[$\; s = 2$, $T= 0.85$.\hspace{0 cm}] {\includegraphics[width=8cm]{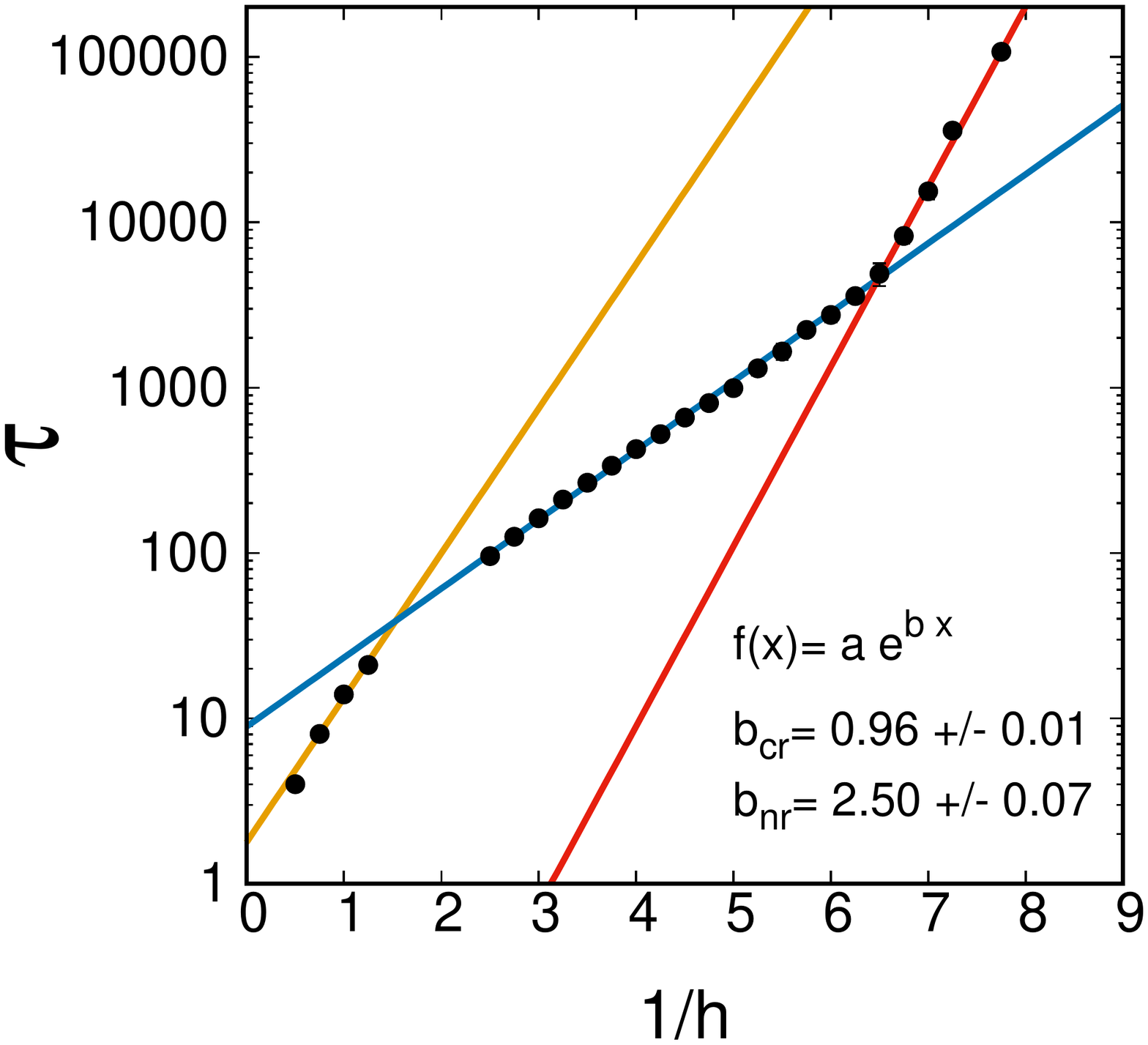}}\\
\subfigure[$\; s = 5/2$, $T= 0.79$.\hspace{0 cm}] {\includegraphics[width=8cm]{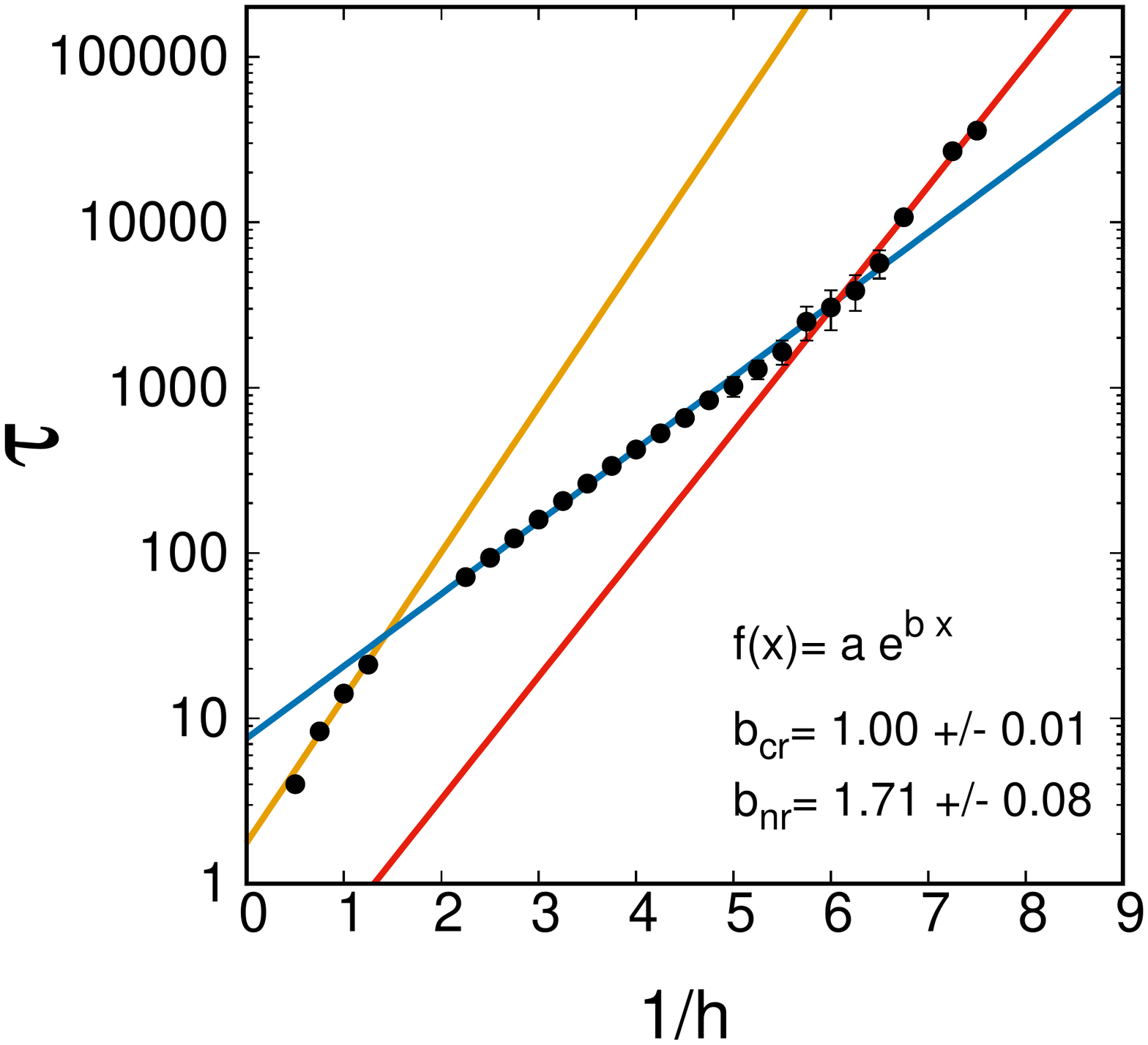}}
\subfigure[$\; s = 3$, $T= 0.76$.\hspace{0 cm}] 
{\includegraphics[width=8cm]{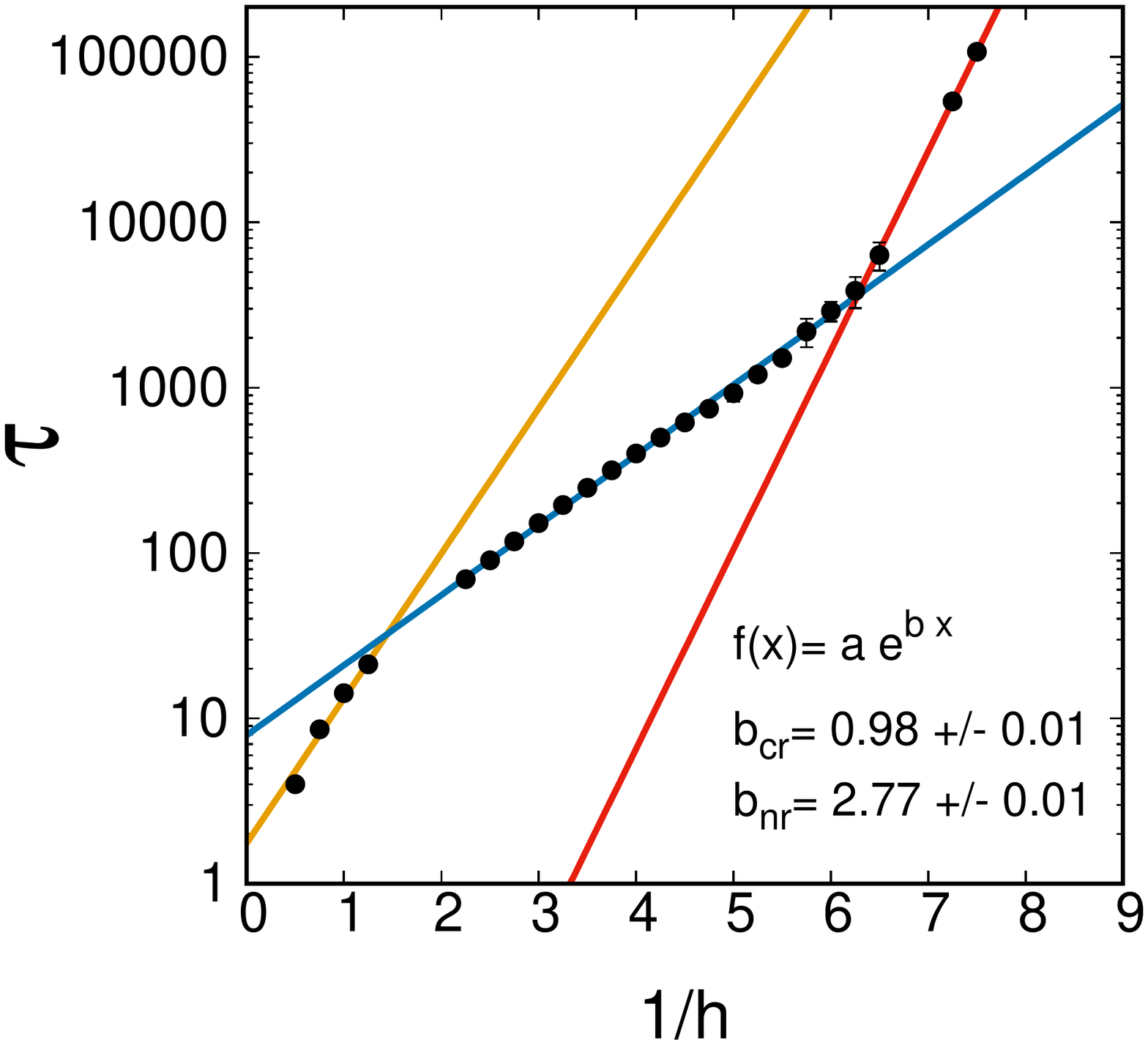}}

\caption{Same as in Fig.~\ref{isingbeck} for four spin-$s$ Blume-Capel models. Again, in some of the data points error bars are smaller than the symbol size. See also Tab.~\ref{chisquare-bc}.}\label{bcbeck}
\end{figure}

An analogous illustration is presented in Fig.~\ref{bcbeck} for the anisotropic spin-$s$ Blume-Capel model with $s= 3/2, 2, 5/2$, and $3$. The obtained results indicate that the Becker-D\"{o}ring analysis of three different reversal 
spin mechanisms is again verified. One minor comment here is that for a particular $s$-value, the reversal time of spin-$s$ Blume-Capel model is found to decrease in comparison to the Ising case. Actually, due to the presence of anisotropy some of the states will be energetically more favorable so that the effective bulk energy in Eq.~(\ref{eq:2}) will be increased, causing the decrease in $l_{\rm c}$ as well in the metastable lifetime. This phenomenon has also been reported very recently in Ref.~\cite{naskar2}. Table~\ref{chisquare-bc} gives a descriptive summary of our statistical tests corresponding to Fig.~\ref{bcbeck}, again in favor of our approach. 

\begin{figure}[h!]
\subfigure[$\; h=-0.2,\; \tau= 1915 \pm 291$ MCSS. \hspace{0 cm}] {\includegraphics[width=8cm]{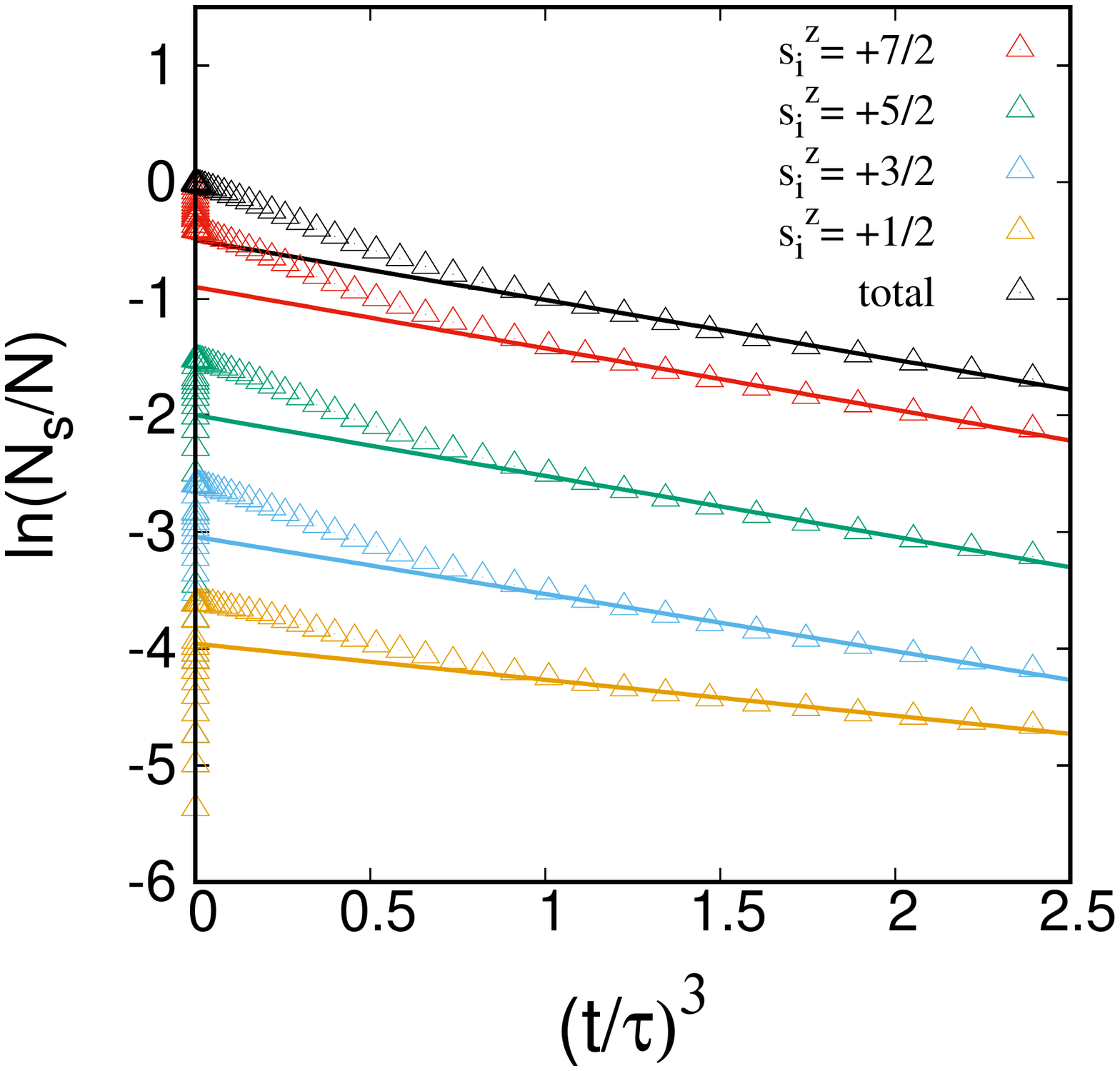}}
\subfigure[$\; h=-0.5,\; \tau= 74 \pm 1 $ MCSS.\hspace{0 cm}] {\includegraphics[width=8cm]{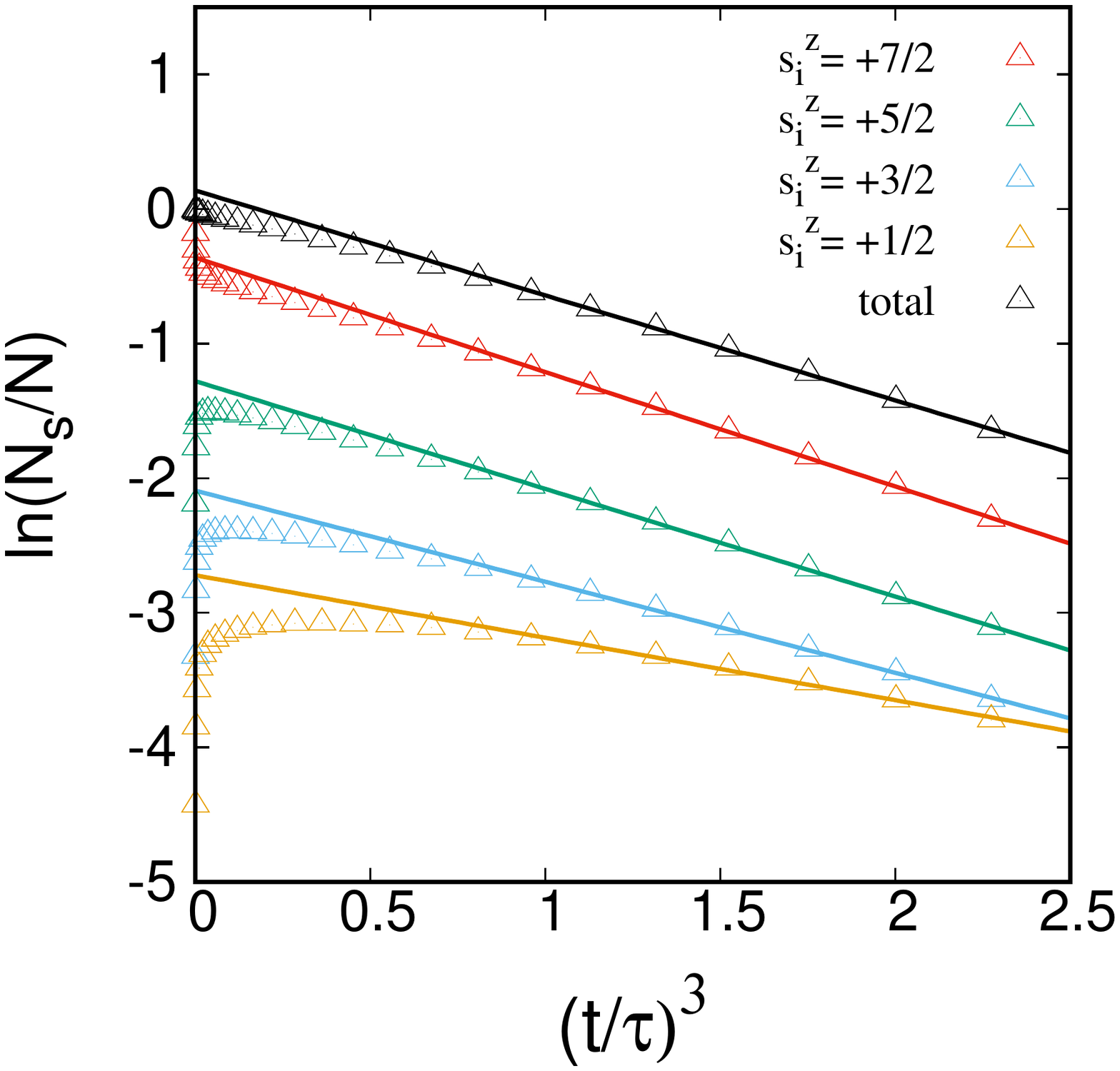}}\\
\subfigure[$\; h=-0.8,\; \tau= 27 \pm 1$ MCSS.\hspace{0 cm}] {\includegraphics[width=8cm]{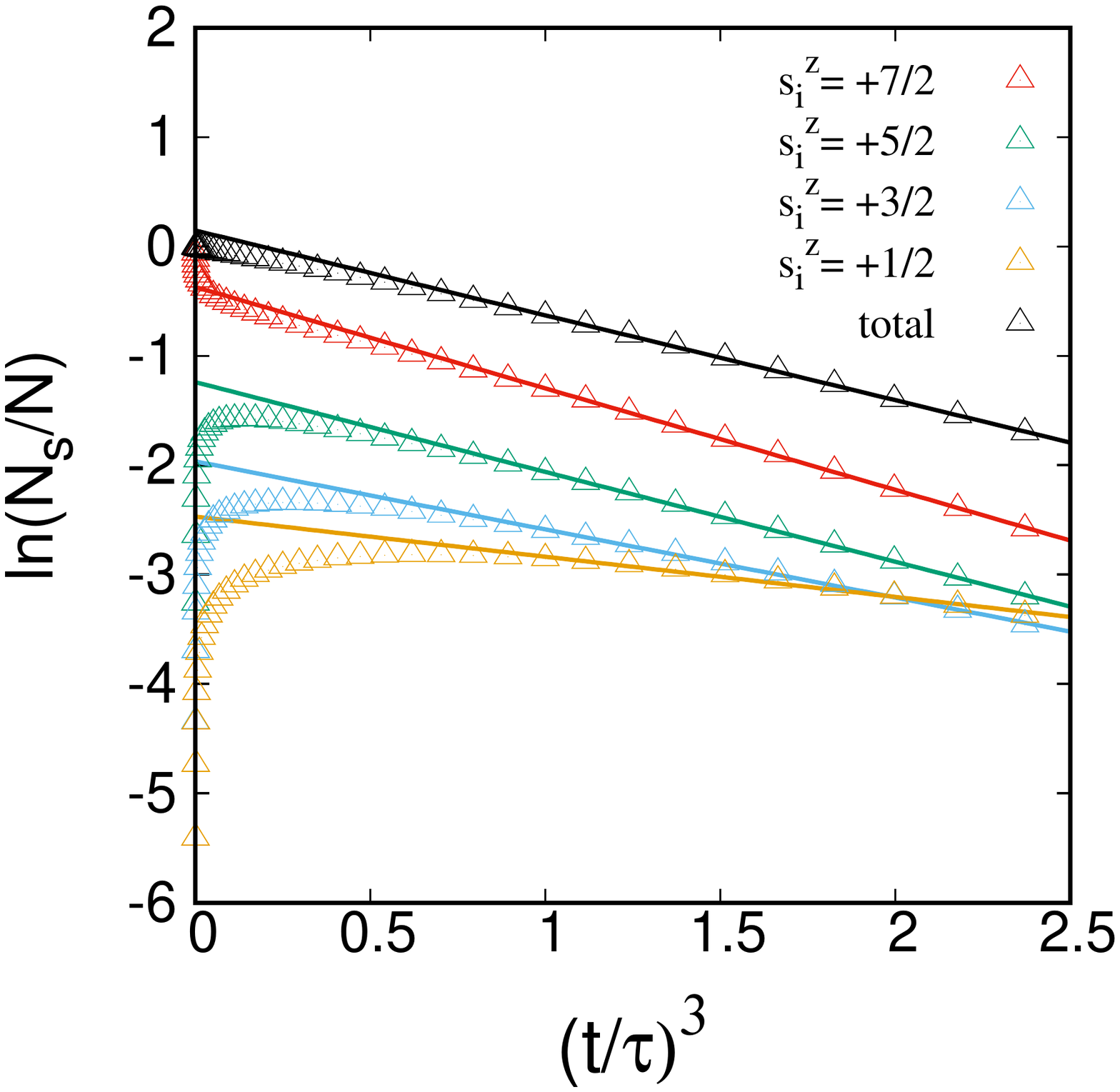}}

\caption{Variation of $\ln{(N_{\rm s}/N)}$ of the spin components $s_i^z = \{7/2, 5/2, 3/2, 1/2\}$ and also of the total versus $(t/\tau)^{3}$ for the $s = 7/2$ Ising system in the presence of three different strengths of the applied field. $N_{\rm s}$ is the number of particular spin state $s_i^z$, $N$ is the total number of spins, and $\tau$ the reversal time. For the case of black open triangles, $N_{\rm s}$ is the total number of the four spin components. The temperature is set to $T = 0.7T_{L}^{\ast} $ and the fits shown by the solid lines correspond to times $t > \tau$. Results averaged over $1000$ samples are shown with the size of error bars being smaller than that of the symbol size used.}\label{isingavrami}
\end{figure}

\begin{figure}[h!]
\subfigure[$\; h=-0.2,\; \tau= 1019 \pm 139$ MCSS. \hspace{0 cm}] {\includegraphics[width=8cm]{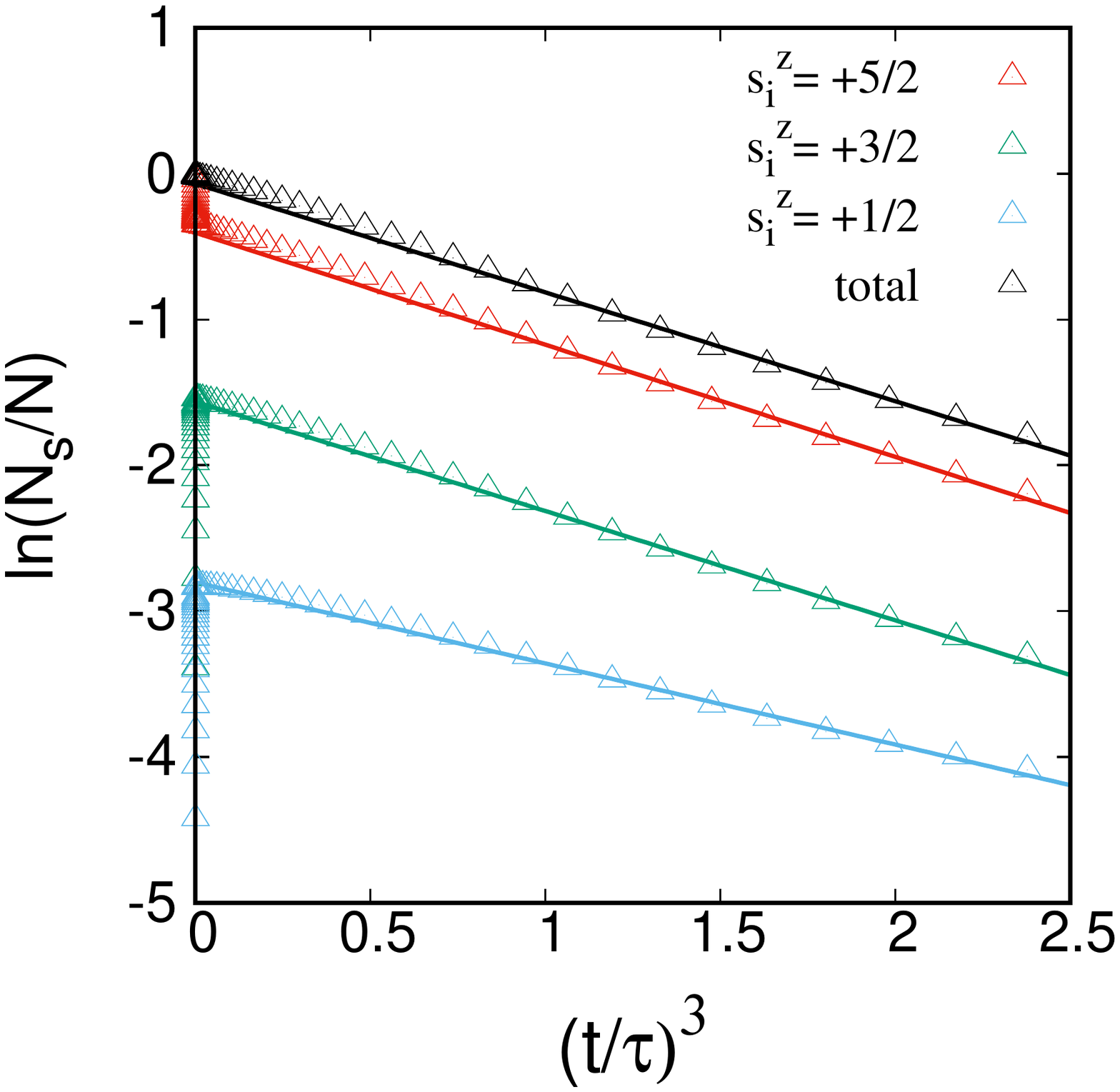}}
\subfigure[$\; h=-0.5,\; \tau= 54 \pm 1$ MCSS. \hspace{0 cm}] {\includegraphics[width=8cm]{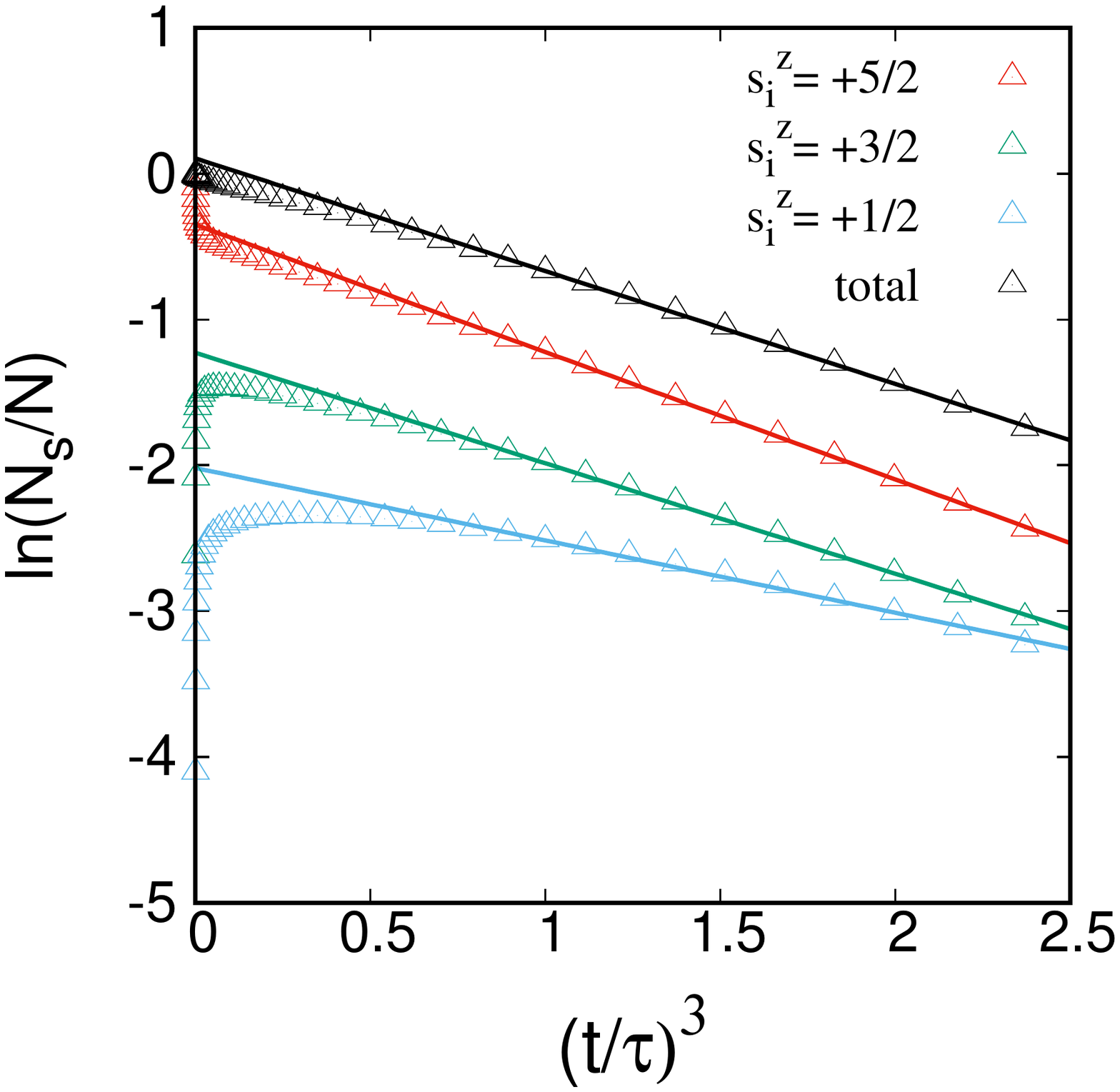}}\\
\subfigure[$\; h=-0.8,\; \tau= 23 \pm 1$ MCSS. \hspace{0 cm}] {\includegraphics[width=8cm]{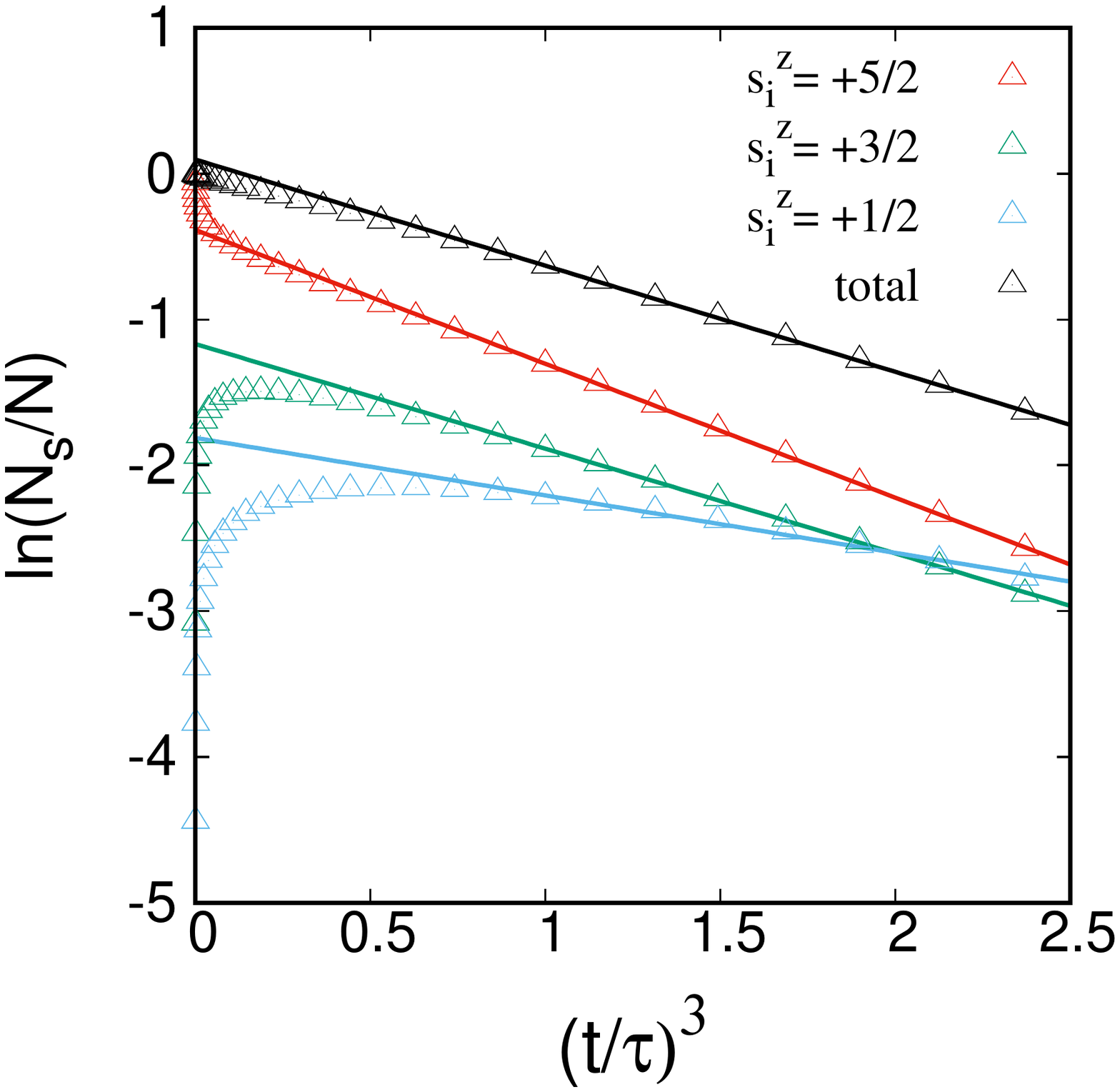}}

\caption{Similar to Fig.~\ref{isingavrami} for the spin components $s_i^z= \{5/2,3/2,1/2\}$ and also for the total of them for the $s = 5/2$ Blume-Capel model. Again, as in Fig.~\ref{isingavrami} the size of error bars is smaller than the symbol size.}\label{bcavrami}
\end{figure}

At this stage, the time dependence of the metastable's volume fraction decay is crucial in order to obtain an idea about the spin dynamics of our problem. According to Avrami's law~\cite{avrami} the logarithm of the metastable volume fraction in a $d$-dimensional Ising system decays as $\sim t^{d+1}$ close to $T_{\rm c}$. This behavior has already been verified in the two-dimensional random-field Ising model~\cite{naskar1} and the spin-$1$ Blume-Capel model~\cite{naskar2}. Here we have checked the validity of this law for the spin-$7/2$ Ising system at $T=0.7T_{L}^{\ast}$, see Fig.~\ref{isingavrami}. We considered the relative abundance of the $7/2$, $5/2$, $3/2$, and $1/2$ states as the metastable volume fraction and also their total relative abundance is presented (black line in Fig.~\ref{isingavrami}). The results indicate that the Avrami's law in the spin-$7/2$ Ising system is well obeyed. In addition, under the influence of a weak field the logarithm of the volume fraction decays faster in the vicinity of the metastable lifetime $(t/\tau)^3=1$ compared to latter times.  On the other hand, in the presence of strong fields the decay rate is higher at $t>\tau$. Additionally, a peak at the initial stage is observed for the density 
of $s_i^z= 5/2, 3/2$, and $1/2$. This should be expected because initially 
the $7/2$ state will decay into those states which will then decay to other 
subsequent ones, \emph{i.e.}, $s_i^z= -1/2, -3/2$, and $-5/2$. In full analogy to Fig.~\ref{isingavrami}, Fig.~\ref{bcavrami} is a manifestation of the Avrami's law for the anisotropic spin-$5/2$ Blume-Capel model.

\section{Summary}
\label{summary} 

In summary, we have investigated the metastable behavior of general discrete spin-$s$ models. Classical nucleation theory, as proposed by Becker and D{\"o}ring many years ago, has been nicely verified in a number of recent studies for the spin-$1/2$ Ising ferromagnet. However, up to this day it remained unclear whether the theory accommodates for metastability phenomena of general spin-$s$ ferromagnetic models. Other open questions under scrutiny include the effect of single-ion anisotropy on the reversal time of the magnetization and decay of metastable volume fraction. We believe that these questions have been satisfactorily addressed in the current paper.

In particular our Monte Carlo results for the general Ising and Blume-Capel models fully verified predictions of the Becker-D{\"oring} theory, indicating its wide applicability to the general case of spin-$s$ ferromagnetic models. Redefining the metastable volume as the relative abundance of the positive values of the spin projections, we studied its variation as a function of the time for three different regimes of the applied magnetic field (weak, intermediate, and strong). In all cases, the metastable volume fraction was found to decay following the Avrami's law. Finally, many of the intriguing phenomena revealed via Monte Carlo simulations were explained using simple energetic arguments of droplet formation in the underlying theory. 

Altogether, as it became apparent from this work the system becomes more flexible for reversal as the value of $s$ increases. At this stage it would be very interesting to investigate the reversal mechanisms for continuous spin Ising and Blume-Capel models, possibly determining the limit of flexibility in these reversal processes. In any case the technological importance of using magnets with higher spin values  for magnetic recording and switching has been highlighted very recently in Ref.~\cite{shankar} and we hope that this work will pave the way for more advanced studies in this growing field of research. Noteworthy, the Blume-Capel model considered in this work has been widely used up to now in the literature for modeling various physical systems with experimental analogues in and out of equilibrium, see Ref.~\cite{fytas20} and references therein. Yet other fruitful candidates would be the general $q$-states Potts model and the anisotropic Heisenberg
ferromagnet (having continuous spin symmetry) where compelling metastability results are expected while varying the number of spin states $q$ and anisotropy. We plan to pursue these research lines in the near future.

\begin{acknowledgments}
We would like to thank the two anonymous referees who have helped us improve our manuscript with their instructive comments. M.~N. would like to thank the Swami Vivekananda Scholarship (SVMCMS) for financial  support. M.~A. acknowledges financial support from the FRPDF grant of Presidency University. We would also like to thank Ms. A. Mukherjee for her suggestion in common online Latex editing. The numerical calculations reported in this paper were performed at T\"{U}B\.{I}TAK ULAKBIM, High Performance and Grid Computing Center (TR-Grid e-Infrastructure).
\end{acknowledgments}

\newpage

\begin{table}[ht!] 
	\caption{Pseudocritical temperatures of the $L = 100$ spin-$s$ Ising and Blume-Capel models obtained from the peak location of the magnetic susceptibility. A uniform maximum error of $10^{-2}$ stems from the temperature-step $\delta T = 0.01$ in our simulations. }
	\centering
	\begin{tabular}{ c|c|c }
		\hline\hline
		Spin-$s$ systems & Ising models & Blume-Capel models\\
		 $s$ & $T_{L}^{\ast}(\Delta=0)$ & $T_{L}^{\ast}(\Delta=0.5)$ \\  \hline	
		1/2 & 2.27(1) & (not applicable) \\		
		1 & 1.72(1) & 1.58(1)\\			
		3/2 & 1.47(1) & 1.33(1)\\		
		2 & 1.34(1) & 1.21(1)\\  		
		5/2 & 1.26(1) & 1.13(1)\\	
		3 & 1.21(1) & 1.08(1)\\			
		7/2 & 1.18(1) & 1.04(1)\\
		\hline \hline		
	\end{tabular}
\label{tctable}
\end{table}


\begin{table}[ht!] 
	\caption{Fitting parameters corresponding to Fig.~\ref{isingbeck}. We have used the standard $\chi^2/{\rm DOF}$ test for goodness of fit, where DOF denotes the number of degrees of freedom. Specifically, the $Q$ value of our $\chi^2$ tests is the probability of finding a value of $\chi^2$ larger than the one actually found from our numerical data -- see also the relevant discussion in the main text.}
	\centering
	\begin{tabular}{|c|*{7}{c|}}
		\hline
		\multirow{2}{*}{Spin-$s$ systems} & \multicolumn{3}{c|}{Coalescence regime} & \multicolumn{3}{c|}{Nucleation regime}\\
		\cline{2-7}  & $\chi^{2}$ & DOF & $Q$ & $\chi^{2}$ & DOF & $Q$\\	
		\hline
		1/2 & 6.5798 & 7 & 0.4739 & 3.7075 & 2 & 0.1566\\	\hline
		2 & 17.0516 & 12 & 0.1477 & 3.4113 & 2 & 0.1816\\	\hline
		5/2 & 13.5384 & 12 & 0.3311 & 4.9426 & 3 & 0.1761\\	\hline
		3 & 16.5078 & 13 & 0.2228 & 0.1689 & 1 & 0.6811\\	\hline
	\end{tabular}
	\label{chisquare-is}
\end{table}


\begin{table}[ht!] 
	\caption{Fitting parameters corresponding to Fig.~\ref{bcbeck}.}
	\centering
	\begin{tabular}{|c|*{7}{c|}}
		\hline
		\multirow{2}{*}{Spin-$s$ systems} & \multicolumn{3}{c|}{Coalescence regime} & \multicolumn{3}{c|}{Nucleation regime}\\
		\cline{2-7}  & $\chi^{2}$ & DOF & $Q$ & $\chi^{2}$ & DOF & $Q$\\	
		\hline
		3/2 & 11.3564 & 13 & 0.581 & 3.9435 & 3 & 0.2676\\	\hline
		2 & 11.322 & 14 & 0.6606 & 3.9403 & 3 & 0.268\\	\hline
		5/2 & 13.3257 & 14 & 0.5010 & 5.5415 & 3 & 0.1362\\	\hline
		3 & 16.3331 & 14 & 0.2934 & 0.4569 & 2 & 0.7958\\	\hline
	\end{tabular}
	\label{chisquare-bc}
\end{table}


\end{document}